\documentclass[prd,a4paper,aps,showpacs,twocolumn,floats]{revtex4}
\usepackage{epsfig}
\usepackage{graphicx,amsmath,amsfonts,amssymb,slashed,upgreek,color,caption}

\newcommand{\bk}{{\mathbf k}}
\newcommand{\boe}{{\mathbf e}}

\newcommand{\bq}{{\mathbf q}}

\newcommand{\bB}{{\mathbf B}}
\newcommand{\bE}{{\mathbf E}}

\newcommand{\bx}{{\mathbf x}}

\newcommand{\al}{\alpha}

\newcommand{\de}{\delta}
\newcommand{\De}{\Delta}
\newcommand{\ep}{\epsilon}
\newcommand{\ga}{\gamma}

\newcommand{\la}{\lambda}
\newcommand{\Om}{\Omega}

\newcommand{\si}{\sigma}

\newcommand{\vph}{\varphi}
\newcommand{\ra}{\rightarrow}

\newcommand{\be}{\begin{equation}}
\newcommand{\ee}{\end{equation}}

\newcommand{\bea}{\begin{eqnarray}}
\newcommand{\eea}{\end{eqnarray}}
\newcommand{\bean}{\begin{eqnarray*}}
\newcommand{\eean}{\end{eqnarray*}}

\newcommand{\HH}{{\cal H}}
\newcommand{\A}{{\mathcal A}}

\newcommand{\ik}{\frac{i k^j}{ak^2}}
\newcommand{\qj}{q_{\eb\, j} }

\newcommand{\eb}{{\rm em}}
\newcommand{\pis}{{\Pi_S}}
\newcommand{\pisp}{{\Pi'_S}}

\newcommand{\hpis}{{\Omega^{-}_\Pi}}
\newcommand{\hpisp}{{\Omega^{+}_\Pi}}

\newcommand{\SB}{{S_{\rm em}}}

\begin{document}

\title{Magnetic fields from inflation: the transition to the radiation era}

\author{Camille Bonvin$^{1}$, Chiara Caprini$^{2}$ and Ruth Durrer$^{3}$
\\  }
\affiliation{$^{1}$ Kavli Institute for Cosmology Cambridge and Institute of Astronomy,
Madingley Road, Cambridge CB3 OHA, UK\\
and\\ DAMTP, Centre for Mathematical Sciences, Wilberforce Road, Cambridge CB3 OWA, UK\\
${}^{2}$CEA, IPhT and CNRS, URA 2306, F-91191 Gif-sur-Yvette, France \\
${}^{3}$D\'epartement de Physique Th\'eorique and Center for Astroparticle Physics, Universit\'e de 
Gen\`eve, 24 quai Ernest 
Ansermet,~CH--1211 Gen\`eve 4, Switzerland}

\date{\today}

\begin{abstract}
We compute the contribution to the scalar metric perturbations from large-scale magnetic fields which 
are generated during inflation.
We show that apart from the usual passive and compensated modes, the magnetic fields also contribute
to the constant mode from inflation. This is different from the causal (post-inflationary) generation
of magnetic fields where such a mode is absent and it might lead to significant, non-Gaussian 
CMB anisotropies.
\end{abstract}


\maketitle

\section{Introduction}
\label{sec:intro}

Magnetic fields are observed in cosmic structures over a wide range of length-scales and redshifts; from 
galaxies to regions around high redshift quasars, from clusters and superclusters to low density filamentary 
regions~\cite{filaments}. The field values are a few microgauss in galaxies and clusters, and of the order of 
the nanogauss in filaments. Recently, Fermi and HESS data have been used to put a lower bound of at least 
$10^{-17}$G on the intensity of magnetic fields in the intergalactic medium and even in voids~\cite{voids}. Finding an explanation for these magnetic fields is challenging, and their origin to date remains an open problem. 

One possibility is that magnetic fields have been generated in the primordial Universe \cite{grasso}. In particular, primordial magnetogenesis mechanisms operating during inflation have the advantage to provide magnetic seeds filling the entire Universe, possibly with significant amplitude also at very large scales. This goes in the right direction to explain both the ubiquity of the observed fields and the uniformity of the measured amplitudes. 

In this paper, we focus on inflationary magnetic fields, generated by breaking conformal invariance of electromagnetism via a term in the action of the form $f^2(\varphi)F^{\mu \nu}F_{\mu \nu}$. This coupling of the electromagnetic  field to the inflaton was first proposed in Refs.~\cite{turner,ratra}, and subsequently reanalyzed in~\cite{yokoyama} for different categories of string-inspired inflationary scenarios (see also \cite{Campanelli:2008qp}). The same kind of action has been considered in the context of dilaton electromagnetism \cite{bamba1} and DBI (Dirac-Born-Infeld) inflation \cite{bamba2}. 
In Ref.~\cite{demozzi} it has been pointed out that, since $f(\varphi)$ plays the role of the inverse coupling 
constant to the charged Dirac field (e.g. the electron), it must remain large because perturbation theory of
the interaction of the Dirac field with the electromagnetic field is only trustable in the small coupling regime. 
This constraint greatly reduces the capability of the model to give rise to a significant magnetic field amplitude. Recently, it has been pointed out that this problem could be circumvented by coupling not only the electromagnetic field but the entire matter Lagrangian to the inflaton~\cite{1109.4415}. However, a multiplication of the entire matter Langrangian 
with $f^2(\varphi)$ could be absorbed in a field redefinition and will therefore not lead to any physical effects. Another possibility to circumvent the problem would be to multiply the coupling term with $f(\varphi)$, i.e. 
$\bar\psi e\ga^\mu A_\mu\psi \ra \bar\psi ef(\varphi)\ga^\mu A_\mu\psi $. This ensures that the coupling to the canonically normalized field, $f(\varphi)A_\mu$ remains constant, however this coupling explicitly breaks gauge invariance which is only recovered 
when $f(\varphi)$ freezes in after inflation. 

In this work we still concentrate on a coupling of the form $f^2(\varphi)F^{\mu \nu}F_{\mu \nu}$, because it has the advantage to be quite general and simple, and because we believe that the main feature of the result we obtain does not depend on the specific form of the coupling. Note that helical magnetic fields can be generated by coupling a pseudoscalar inflaton to the $\widetilde F F$-term. However, it has been shown that this coupling generically leads to blue spectra, which do not have enough power on large scales to be the
seeds of the large-scale coherent fields observed in galaxies and clusters~\cite{Durrer:2010mq}.
 
Starting from the consistent assumption that the electromagnetic field arising from the amplification of vacuum fluctuations is subdominant and does not affect the background dynamics of inflation, previous analyses have evaluated the spectrum of the electromagnetic energy density and studied the conditions under which this kind of coupling gives rise to interesting magnetic field amplitudes after inflation~\cite{ratra,yokoyama}. However, the contribution of the electromagnetic energy density, even if it does not affect the background, still affects metric perturbations at first order in perturbation theory. Therefore, here we proceed one step further and calculate the scalar metric perturbations induced by the electromagnetic field at first order in perturbation theory on superhorizon scales. We assume that the electromagnetic energy-momentum tensor is first order, and consequently the electromagnetic field is half order. Another
possibility would be to set the electromagnetic field first order, and its energy-momentum tensor
second order. However, the metric perturbations induced by the electromagnetic field turn out 
to be much larger than second-order perturbations in the inflaton; the latter can therefore be consistently neglected, effectively going back to the first-order scheme. The point of view of treating the electromagnetic energy-momentum tensor as a second-order perturbation has been taken in Ref.~\cite{barnaby} where, nevertheless, the second-order
inflaton perturbations have also been neglected: consequently,  Ref.~\cite{barnaby} effectively adopts 
the perturbation expansion used in this work.

In the present analysis we find that during inflation, the large-scale solution for the Bardeen potentials is 
$\Psi_- \,,~ \Phi_- \sim \hpis (k,\eta)/(k\eta)^2$, sourced by $\hpis$ which denotes the electromagnetic anisotropic stress normalized to the background energy density (note that here we identify the metric perturbation with their r.m.s. (root of the mean square) amplitude, c.f. discussion in Section \ref{sec:bardeeninflation}). Therefore, even though $\hpis \ll 1$, the Bardeen potentials become large on superhorizon scales. However, we demonstrate that the ratio of the Weyl to Ricci tensors is small, because it is determined by $\hpis \ll 1$. It also turns out that perturbations in other gauges, like comoving and synchronous gauge do remain small: therefore perturbation theory is valid during inflation.

We then match the inflationary solution for the metric perturbations at the end 
of inflation to the solution in the radiation era on the surface of constant background energy density, in 
the usual way \cite{muk_deruelle}.  From this, we derive the metric perturbations at superhorizon scales during the radiation-dominated era. The matching shows that the large, $1/(k\eta_*)^2$ contribution to the Bardeen potential is transferred entirely 
to the decaying mode in the radiation era. On the other hand, $\Psi_+$ during the radiation era gets a contribution at next order in the large-scale expansion, i.e. at order $\mathcal{O}((k\eta_*)^0)$. 
This constant term adds to the usual `passive' and `compensated' modes leading to a new effect in cosmic microwave background (CMB) anisotropies, and it is specific to inflationary generated magnetic fields (it is absent if the magnetic field is generated by a causal process). 

The rest of the paper is organized as follows: in Section \ref{sec:metric} we present the  perturbed Einstein equations and derive the  Bardeen equation in the presence of a nonzero electromagnetic source; in Section \ref{sec:source} we assume a power-law evolution in time for $f(\varphi)$, and  calculate the scalar electromagnetic anisotropic stress which represents the dominant source term of the Bardeen equation at superhorizon scales; in Section \ref{sec:solution}, we find the solutions for the Bardeen potentials both in the inflationary and radiation-dominated eras, and perform the matching. In Section \ref{sec:curvature} we derive the solutions at next-to-leading order in the large-scale expansion and in Section \ref{sec:con} we conclude. Some details of the calculations are deferred to appendices.

{\bf Notation:} Throughout this paper we use conformal time $\eta$, comoving space coordinates ${\bf x}$ and wave vectors ${\bf k}$ with the metric $ds^2=a^2(\eta)(-d\eta^2+ \de_{ij}dx^i dx^j)$; 4d spacetime indices are Greek letters while 3d spatial indices are Latin letters and spatial vectors are denoted in bold face. For the metric and scalar field perturbations we follow the conventions of \cite{mukhanov}, while for the electromagnetic action and field quantization we follow the conventions of \cite{yokoyama,subramanian}.  We define the Planck mass by  $m_P = (\sqrt{8\pi G})^{-1}$.

\section{Metric perturbations sourced by the electromagnetic field}
\label{sec:metric}

We consider an electromagnetic field generated during inflation by breaking of conformal invariance, as specified in Section \ref{sec:source}. 
We assume that the electromagnetic energy-momentum tensor is first order in perturbation theory,
meaning that the electric and magnetic fields are half order (c.f. discussion in the Introduction). 
During inflation, the background evolution is therefore determined only by the background scalar field, 
whereas both the scalar field perturbation and the electromagnetic field contribute to the first
order energy-momentum tensor, ${\delta T^\al}_\beta={{\de T_\varphi}^\al}_\beta+T^\al_{\eb\, \beta}$.
The scalar field driving inflation is split into a background part and a first-order perturbation as
$\varphi(\bx,\eta)=\varphi_0(\eta)+\de\varphi(\bx,\eta)$, so that the background equations are \cite{mukhanov}
\be
4\pi G (\varphi_0')^2=\HH^2-\HH' \quad \mbox{and} \quad
\varphi_0''+2\HH\varphi_0'+a^2V_{,\varphi}=0\; .
\label{back}
\ee
A prime denotes derivative with respect to conformal time $\eta$ and $\HH=a'/a$.
Since the electromagnetic field is half order, the electromagnetic energy-momentum tensor can be decomposed with respect to the unperturbed velocity of the FL background, $\bar{u}^\al =a^{-1}(1,{\bf 0})$, and to the unperturbed metric $\bar{g}_{\al \beta}$, and it is gauge invariant \cite{cc}:
\be
T^{\al\beta}_{\eb}=(\rho_\eb +p_\eb)\bar{u}^\al\bar{u}^\beta +p_\eb\bar{g}^{\al \beta}
+2 \bar{u}^{(\al}q^{ \beta)}_{\eb} +\Pi^{\alpha\beta}_{\eb}\; . 
\label{Tem}
\ee
The parentheses around the superscripts of the third term indicate symmetrization.
We focus on scalar perturbations and use longitudinal gauge with the notation~\cite{mukhanov}
\be
ds^2 = a^2\left[-(1+2\Phi)d\eta^2 + (1-2\Psi)d\bx^2\right] \,.
\ee
Note that the `names' $\Phi$ and $\Psi$ are interchanged with respect to~\cite{mybook}.
The perturbed Einstein equations, $\delta G^\al_{\ \beta}=8\pi G\delta T^{\al}_{\ \beta}$, 
for scalar perturbations in Fourier space in the presence of an electromagnetic field are:
 \begin{align}
&3\HH\Psi'+(2\HH^2+\HH')\Phi+k^2\Psi=\nonumber\\
&-4\pi G\Big(\varphi_0'\delta\varphi'
+V_{,\varphi}a^2\delta\varphi\Big)-4\pi Ga^2\rho_\eb\; ,\\
&\Psi''+2\HH\Psi'+\HH\Phi'+(2\HH^2+\HH')\Phi-\frac{k^2}{3}(\Phi-\Psi)=\nonumber\\
&4\pi G\Big(\varphi_0'\delta\varphi'
-V_{,\varphi}a^2\delta\varphi\Big)+4\pi Ga^2p_\eb\; ,\\
&\Psi'+\HH\Phi=4\pi G\varphi_0'\delta\varphi-4\pi Ga\, {\rm i} \, \frac{k^j}{k^2}\, q_{\eb\, j}\; ,\\
&k^2(\Phi-\Psi)=-8\pi Ga^2\pis\; . \label{phiminuspsi}
\end{align}
Here $\rho_\eb(\bk)$, $p_\eb(\bk)$ and $q_{\eb\, j}(\bk)$ are the electromagnetic field energy density, pressure and Poynting vector in Fourier space obtained from Eq.~\eqref{Tem}.  $\pis(\bk)\equiv -3/2 \hat k^i\hat k_j \Pi^{\ j}_{\eb\, i} (\bk)$ is the scalar part of the electromagnetic anisotropic stress; it is of the same order of magnitude as
the electromagnetic energy density.
These equations can be combined into a second-order evolution equation for the variable $\Psi$, the Bardeen equation,
\be
\label{psi}
\Psi''+2\left(\HH-\frac{\varphi_0''}{\varphi_0'}\right)\Psi'+
\left(2\HH'-\frac{2\HH\varphi_0''}{\varphi_0'}+k^2\right)\Psi=S \; .
\ee
The source term $S$ is due to the presence of the electromagnetic field. It is given by
\be
\label{source}
\begin{split}
&S=8\pi Ga^2\Bigg[\HH\frac{(a^2\pis)'}{a^2k^2}
+2\left(\HH'-\HH\frac{\varphi_0''}{\varphi_0'}\right)\frac{\pis}{k^2}-\frac{\pis}{3}\\
& -\frac{1}{2}(\rho_\eb-p_\eb)+\left(2\HH+\frac{\varphi_0''}{\varphi_0'}\right)\frac{ {\rm i}\, k^jq_{\eb\, j}}{k^2 a} \Bigg]\; .
\end{split}
\ee
We want to solve Eq.~(\ref{psi}), in order to determine the effect of the electromagnetic field on the scalar metric perturbations. We are interested in the solution at very large scales $k|\eta|\ll 1$. We consider slow-roll inflation with
\be
 a\simeq a_1 \left|\frac{\eta_1}{\eta}\right|^{1+\epsilon} ~,\quad \HH\simeq -\frac{1+\epsilon}{\eta}  
\label{aHdef}
\ee
at first order in slow-roll. The slow-roll parameters $\ep$ and $\ep_2$ are defined by~\cite{mybook}
\be\label{slro}
\quad \HH^2-\HH'=\ep\HH^2 ~, \qquad \ep'=2\ep(3\ep_2+2\ep)\HH \,.
\ee
From this we infer
\be
\frac{ \varphi''_0}{\varphi'_0} -\HH = \left(3\ep_2+\ep\right)\HH \,. \label{slro2}
\ee
Using these expressions together with Eqs.~\eqref{back}, and defining the new variable $x=|k\eta| =-k\eta$, Eq.~\eqref{psi} can be rewritten as 
\be
\label{psiroll}
\frac{d^2\Psi}{dx^2}+\frac{2(\ep+3\ep_2)}{x}\frac{d\Psi}{dx}+\bigg[1-\frac{2(3\ep_2+2\ep)}{x^2}\bigg]\Psi=\frac{S}{k^2}\; .
\ee
At very large scales $x\ll 1$ and at first order in the slow-roll parameters, the source term in \eqref{source} setting $\SB=S/k^2$ reduces to 
\bea
\SB &\simeq & \frac{8\pi Ga^2}{k^4}\Bigg[\HH\pis'
+2\left(\HH^2+\HH'-\HH\frac{\varphi_0''}{\varphi_0'}\right)\pis\Bigg]  \nonumber \\  
&\simeq & \frac{3}{x^2\rho_\varphi}\left[ \frac{2(1+2\epsilon-3\epsilon_2)}{x^2} \pis -\frac{1+3\epsilon}{x}
  \frac{d\pis}{dx}\right]   \label{sourceLS}
\eea
All other contributions to the source term are suppressed by at least one factor $x\ll 1$. The source 
is therefore completely dominated by the electromagnetic anisotropic stress $\pis$. We evaluate $\pis$ in the next section; for this, we have to specify the generation mechanism for the electromagnetic field which is operating during inflation. The above expression for the source includes also terms at first order in the slow-roll expansion: as will become clear in the following (c.f. Section \ref{sec:solution}), for the problem at hand it is not enough to solve the Bardeen equation at lowest order in the slow-roll expansion, but we will need to go to first order. 

In the following, we will also need to solve for the curvature perturbation (see also~\cite{suyama}). The curvature on comoving hypersurfaces $\zeta$ is defined by
\be
\label{zeta}
\zeta=\Psi+2(\HH\Phi+\Psi')/[3\HH(1+w)]\;.
\ee
This variable has the advantage that it is known to be constant on superhorizon scales 
if the source is absent. Using the definition above, we can derive a first-order equation for $\zeta$ which shows that even in the presence of the electromagnetic source, $\zeta$ is conserved at lowest order in the large-scale expansion 
$x\ll 1$, i.e., it does not contain a $1/x^2$ term. Deriving \eqref{zeta}, and eliminating $\Phi$ via 
Eq.~\eqref{phiminuspsi} and $\Psi''$ with the help of the Bardeen Eq.~\eqref{psi}, one obtains: 
\be
\begin{split}
& \zeta' = -  \frac{2 \HH}{3(1+w)} \, \left[  \left(\frac{k}{\HH}\right)^2 \Psi + \frac{\Pi_S}{\rho_\varphi} + \frac{\rho_{\rm em}}{\rho_\varphi} \right.\\ 
& ~~~~~~~~~~~~~~~~~~~~~~~ \left. - \left(2\HH+\frac{\varphi_0''}{\varphi_0'}\right)\frac{3}{\rho_\varphi \, a}  \frac{{\rm i}\, k^j}{k^2}\, q_{\eb\, j}  \right]\,,
\label{zetafirst}
\end{split}
\ee
where $w=p_\varphi /\rho_\varphi =-1+2\ep/3$. 
The curvature perturbation is therefore sourced only at next-to-leading order in the large-scale expansion $x\ll 1$, i.e.
it is of order $x^2\Psi/\ep$. From the above equation, we see that the lowest-order solution for the Bardeen potential 
$\Psi$ is sufficient in order to calculate $\zeta$ at next-to-leading order. 

It is also possible to derive an equation for $\zeta$ which allows us to compute $\zeta$ in a way independent 
of the Bardeen potential. 
This can be done directly from the Einstein equations in comoving gauge; the details of
the derivation are given in Appendix~\ref{app:zeta}, and the resulting equation to lowest order in the slow 
roll parameters is
\be
\label{eq:zeta}
\frac{d^2\zeta}{dx^2} -\frac{2}{x}\frac{d\zeta}{dx}+\zeta =
\frac{1}{\epsilon\, x^2\rho_\varphi}\left[-6\rho_\eb+x\frac{d\rho_\eb}{dx}+x\frac{d\pis}{dx}\right]\,.
\ee
Comparing the source of the Bardeen Eq.~\eqref{sourceLS} and the one of the above equation, we see that
the former is by a factor $x^{-2}$ larger than the latter. We therefore also expect the Bardeen
potentials to be by a factor of about $x^{-2}$ larger than the curvature perturbation, which is only sourced at next-to-leading order in $x\ll 1$. On the other hand, we note that the source of \eqref{eq:zeta} is larger in what concerns the slow-roll expansion: it is of order $\epsilon^{-1}$, while~\eqref{sourceLS} is of order $\epsilon^0$. One therefore needs to be very careful in dealing properly with the large-scale and slow-roll expansions, as will become clear in Section \ref{sec:solution}.  

\section{The source term of the Bardeen equation}
\label{sec:source}

As discussed in the introduction, one of the simplest ways to generate an electromagnetic field by amplification of vacuum fluctuations during inflation is to break conformal invariance of the electromagnetic action by introducing a coupling between the electromagnetic field and the scalar field as
\be
\label{actionem}
S=-\frac{1}{16\pi}\int d^4x \sqrt{-g}f^2(\varphi)F^{\mu \nu}F_{\mu \nu} + \, S_{\varphi,g} + \cdots \; ,
\ee
with the Faraday tensor $F_{\mu\nu}=A_{\nu,\mu}-A_{\mu,\nu}$, and $A_\nu$ the electromagnetic 4-vector potential. In the following, we adopt Coulomb gauge $A_0(\bx,\eta)=0$, $\partial_j A^j(\bx,\eta)=0$ and follow the notation of Ref.~\cite{subramanian}. From Maxwell's equations, $[\sqrt{-g}f^2F^{\mu \nu}]_{,\nu}=0$, we obtain an evolution equation for the space components $A_i(\bx,\eta)$. In a cosmological background it
reads~\cite{subramanian}
\be
A_i''+2\frac{f'}{f}A_i'-\De A_i=0~,
\label{eqA}
\ee
where $\De$ is the comoving spatial Laplacian. For a Fourier mode $k$, we simply have $\De = -k^2$.
The time evolution of the vector potential depends on the coupling function $f(\varphi)$, and we adopt 
the following simple form for it \cite{yokoyama}:
\be
\label{f}
f(\eta)=f_1\left(\frac{\eta}{\eta_1}\right)^\gamma\;. 
\ee 
This choice is motivated on the one hand by simplicity, as it leads to simple power laws for the spectrum of the electromagnetic field. But  it also includes the exponential form first proposed by~\cite{ratra} in the case of power law inflation, and it is sufficiently general to describe the case of generic single-field inflation in the slow-roll approximation. We restrict to the values $-2\leq\gamma\leq 2$, which insures that the electromagnetic field 
remains subdominant and does not back react on the background expansion during 
inflation~\cite{yokoyama,subramanian}. The value $\gamma=-2$ produces a scale-invariant (flat) spectrum for the magnetic field energy density, corresponding to a spectral index $n_B=-3$ for the magnetic field spectrum itself, as defined for example in Eq.~(1.1) of \cite{cc} or Eq.~(\ref{PB}) below. 

With the above time evolution for the function $f$, the equation for the vector potential can be solved analytically. Following~\cite{subramanian} for the quantization of the electromagnetic field, we expand the vector potential in terms of creation and annihilation operators $b_\lambda^\dagger(\bk)$
and $b_\lambda(\bk)$ as
\bea
\label{expansion}
A^i(\bx,\eta)&=&\sqrt{4\pi} \int\frac{d^3k}{(2\pi)^3}\sum_{\lambda=1}^{2} \frac{{\rm e}_\lambda^i(\bk)}{a} \\
&&\Big[b_\lambda(\bk)A(k,\eta) {\rm e}^{i\bk\cdot\bx}
+b_\lambda^\dagger(\bk)A^*(k,\eta){\rm e}^{-i\bk\cdot\bx} \Big]\; , \nonumber
\eea  
where $\boe_1(\bk), ~\boe_2(\bk) $ are unit vectors orthogonal to each other and to $\bk$, which represent the two polarizations of the electromagnetic field. It is convenient to define the new variable $\mathcal{A}=a(\eta)\,f(\eta)\,A(k,\eta)$. Substituting Eq.~\eqref{f} and the expansion Eq.~\eqref{expansion} into Eq.~\eqref{eqA}, this latter can be solved in terms of the variable $\A$ as 
\be
\A(k,\eta)=\sqrt{\frac{x}{k}}\Big[C_1(\ga)J_{\gamma-1/2}(x)+C_2(\ga)J_{-\gamma+1/2}(x)\Big]\; ,
\label{Asolution}
\ee
where $x\equiv |k\eta|=-k\eta$, $J_\nu$ denotes the Bessel function of order $\nu$, and $C_1,~C_2$ are $\ga$ dependent coefficients which are fixed as usual by imposing the initial condition that for subhorizon scales, $-k\eta\rightarrow\infty$, the gauge field is in the Minkowski space vacuum \cite{subramanian}.  

From the above solution for $\A$ we can infer the anisotropic stress $\pis(\bk,\eta)$, which appears in the source term in Eq.~\eqref{sourceLS}. We have $\pis(\bk)=T^{\ i}_{\eb\, i} /2-3/2\hat k^i\hat k_j T^{\ j}_{\eb\, i} $, and the 
electromagnetic energy-momentum tensor is given by
\be
\label{Tmunu}
\begin{split}
& T^{\nu}_{\eb\, \mu}(\bx,\eta)=\frac{f^2}{4\pi} \Big(F_{\mu\al} F^{\nu\al}-\frac{1}{4}\, \bar{g}_{\mu}^{\ \nu}F_{\al\beta}F^{\al\beta} \Big)\;, \\
& T_{\eb\, ij}(\bx,\eta)=\frac{f^2}{4\pi\,a^2} \bigg\{ -A'_i{A'}_j+ (A_{k,i}-A_{i,k}) \\ 
& \hspace{1.2cm} (A_{k,j}-A_{j,k}) -\frac{1}{2}\de_{ij} \left[ (\nabla \wedge {\bf A})^2 - {{\bf A}'}^2 \right] \bigg\}\;.
\end{split}
\ee
Inserting the expansion 
Eq.~\eqref{expansion} into Eq.~\eqref{Tmunu} we find 
\be
\label{pis}
\begin{split}
&\pis(\bk,\eta)=\frac{3}{2a^4}\int\frac{d^3k'}{(2\pi)^3}\!\!\sum_{\la,\la'=1}^{2}\!\!
\left(\frac{\delta^{ij}}{3}-\hat k^i\hat k^j\right)\Bigg\{ \\
&-{\rm e}_{\la\, i}(\bk'){\rm e}_{\la'\, j}(\bk-\bk') f^2 \!\left(\!\frac{\A(k',\eta)}{f} \!\right)'\!\!\!\left(\!\frac{\A^*(|\bk-\bk'|,\eta)}{f} \!\right)'\\
&+ \Big[{\rm e}_{\la'\, \ell}(\bk-\bk')(k_j-k_j')-{\rm e}_{\la'\, j}(\bk-\bk')(k_\ell-k_\ell')\Big]\\
&\times\Big[{\rm e}_{\la\, \ell}(\bk')k_i'-{\rm e}_{\la\, i}(\bk')k_\ell'\Big]\A(k',\eta)\A^*(|\bk-\bk'|,\eta)\Bigg\}\\
&b_\lambda^\dagger(\bk')b_{\lambda'}^\dagger(\bk-\bk')  + \mbox { c.c.}\; .
\end{split}
\ee
Here the `c.c.' stands for the three other terms with operators $b_\lambda(\bk')b_{\lambda'}(-\bk-\bk')$,
$b_\lambda(\bk') b_{\lambda'}^\dagger(\bk+\bk')$ and $b_\lambda^\dagger(\bk')b_{\lambda'}(\bk'-\bk)$. 
More details are given in Appendix~\ref{app:stress}. 

From this expression for $\Pi_S(\bk, \eta)$ we can determine the source 
term in Eq.~\eqref{sourceLS}, which is a quantum operator acting on the electromagnetic vacuum. The source term is of the form
$ S_{\rm em}(\bk,\eta) = \al_1(k,\eta)\pis(\bk,\eta) + \al_2(k,\eta)\pisp(\bk,\eta)$. 
How shall we proceed to compute the induced Bardeen potential? Naively one might simply want to use the
vacuum expectation value of the operator $S_{\rm em}$ as a classical source term. However, this is not sufficient, 
since $\langle 0|\Pi_S|0\rangle$ is independent of position and therefore does not contribute to the fluctuations.
On the other hand, to work with the full fledged operator given in Eq.~\eqref{pis} is a bit unwieldy. The important point to remark, though, is that in order to solve the Bardeen Eq. \eqref{psi} we only need to know the {\it time dependence} of $\Pi_S(\bk, \eta)$, which allows us to calculate also $\pisp(\bk,\eta)$ and therefore the full source term. It turns out that, to determine the time dependence of $\Pi_S(\bk, \eta)$, the easiest way is to first evaluate its power spectrum,
\be
\langle 0| \Pi_S^\dagger(\bq,\eta) \Pi_S(\bk,\eta)|0\rangle = (2\pi)^3P_\Pi(k,\eta)\de(\bq-\bk)\,,
\ee
where the $\de(\bq-\bk)$ is a consequence of translation invariance and the spectrum $P_\Pi(k,\eta)$ depends only on $k=|\bk|$
due to the isotropy of the quantum vacuum.
The details of the calculation are given in Appendix~\ref{app:stress}. Here we only want to stress that, from Eq.~\eqref{pis}, the quantum operator $\Pi_S(\bk, \eta)$ is of the form
\be
\label{e:SOp}
\Pi_S(\bk, \eta)= \sum_{i=1}^4{\bar \Pi}_i(\bk,\eta)O_i(\bk)\,,
\ee
where ${\bar \Pi}_i$ are deterministic functions of time, and the operators $O_i(\bk)$ do not depend on time (we formally perform the integral in $d^3k'$). In the power spectrum $\langle 0| \Pi_S^\dagger(\bq,\eta) \Pi_S(\bk,\eta)|0\rangle  $, only one type of operators $O_i$ is such that
$\langle 0|O^\dagger_i(\bq)O_i(\bk)|0\rangle \neq 0$, namely those which first generate two 
modes and then destroy them. The term which generates and destroys first 
a $\bk'$ and then a $\bq'$-mode only contributes to the zero-mode, not to the fluctuation. There are two terms which give a nonzero contribution, 
and both give the same result. In Appendix~\ref{app:stress} it is shown that finally the anisotropic stress power spectrum can be written as the convolution of the magnetic, electric and Poynting vector power spectra (c.f. Eq.~\eqref{ppiP}):
\begin{widetext}
\bea
\label{ppiP2}
\begin{split}
P_\Pi(k,\eta)=18\pi f^4 \int_0^{1/|\eta|} \frac {k'^2 dk'}{(2\pi)^3}\, 
\Bigg\{& \si_1(\gamma)P_E(k',\eta)P_E(|k-k'|,\eta)
+\si_2(\gamma)P_B(k',\eta)P_B(|k-k'|,\eta) \\
&+ \si_3(\gamma)P_{EB}(k',\eta)P_{EB}(|k-k'|,\eta)\Bigg\} \; .
\end{split}
\eea
\end{widetext}
Here we can neglect the contribution coming from $k'>1/|\eta|$, because for subhorizon modes the 
Bessel functions $J_\nu(k'|\eta|)$ and  $J_\nu(|(k-k')\eta|)$ which enter in the integrand (through Eqs.~\eqref{pis} and \eqref{Asolution}) oscillate, and the result is damped. The prefactors $\si_1(\ga)\,,~\si_2(\ga)$ and $\si_3(\ga)$ depend somewhat on $\gamma$ but are always of order unity, and come from the angular integrals  which cannot be 
evaluated analytically (see Appendix~\ref{app:stress}). The power spectra are defined by 
\bea
\hspace{-0.2cm}\langle 0|B_i(\bq,\eta)B_j^*(\bk,\eta)|0\rangle\!\! &=&\!\!(\de_{ij}-\hat k_i\hat k_j)(2\pi)^3P_B(k,\eta)\de(\bq-\bk) \nonumber\\
\hspace{-0.2cm}\langle 0|E_i(\bq,\eta)E_j^*(\bk,\eta)|0\rangle\!\!&= &\!\!(\de_{ij}-\hat k_i\hat k_j)(2\pi)^3P_E(k,\eta)\de(\bq-\bk) \nonumber \\
\hspace{-0.2cm}\langle 0|E_i(\bq,\eta)B_j^*(\bk,\eta)|0\rangle \!\!&= &\!\!({\rm i}\varepsilon_{ijl}\hat k_l)(2\pi)^3P_{EB}(k,\eta)\de(\bq-\bk) \nonumber
\eea
and have been calculated, e.g., in Refs.~\cite{yokoyama,subramanian} with the results 
\bea
P_B &=& 4\pi \frac{k^2}{f^2a^4} |{\cal A}(k,\eta)|^2\; , \\
P_E &=& 4\pi\frac{1}{a^4} \left|\left(\frac{{\cal A}(k,\eta)}{f}\right)'\right|^2\; ,\\
P_{EB} &=& 4\pi \frac{k}{fa^4} \left(\frac{{\cal A}(k,\eta)}{f}\right)'{\cal A}^*(k,\eta)  \,.
\eea

Since we are interested in the solution for $\Psi$ of Eq.~\eqref{psiroll} at large scales, superhorizon modes, 
 we only need to compute the source for $x=|k\eta| <1$.
We can then expand the Bessel functions in Eq.~\eqref{Asolution} for $x\ll 1$ and in this limit the solution becomes
\bea
\label{Aapp}
\A(k,\eta)&\simeq&\frac{1}{\sqrt{k}}\Big[c_1(\gamma)x^\gamma+d_1(\gamma)x^{\gamma+2} \nonumber \\
&&+\,c_2(\gamma)x^{1-\gamma}+d_2(\gamma)x^{3-\gamma} \Big]\;,
\eea
with
\bea
&& c_1(\gamma)=\frac{{\rm e}^{-{\rm i} \pi \gamma/2}}{\cos(\pi\gamma)}\frac{\sqrt{\pi/4}}{2^{(\gamma-\frac{1}{2})}\Gamma(\gamma+1/2)}\,, \\
&& d_1(\gamma)=\frac{c_1(\gamma)}{\gamma+1/2}\,, \\
&& c_2(\gamma)=\frac{{\rm e}^{{\rm i} \pi (\gamma+1)/2}}{\cos(\pi\gamma)}\frac{\sqrt{\pi/4}}{2^{(\frac{1}{2}-\gamma)}\Gamma(3/2-\gamma)}\,, \\
&& d_2(\gamma)=\frac{c_2(\gamma)}{3/2-\gamma}\,.
\eea
Depending on the value of $\gamma$, different terms dominate in the expansion~\eqref{Aapp}, leading to different results for the magnetic, electric and Poynting vector spectra, and consequently also for $P_\Pi(k,\eta)$.  
For the power spectra we obtain, for $x<1$
\bea
\hspace{-0.2cm}P_B(k,\eta) &= & \frac{4\pi k}{f^2a^4} \left\{\begin{array}{ll} |c_1|^2x^{2\ga} & \mbox{if } \ga<1/2 \\
|c_2|^2x^{2-2\ga} & \mbox{if } \ga>1/2 \,,\end{array} \right.\label{PB}\\
\hspace{-0.2cm}P_E(k,\eta)  &=& \frac{4\pi k}{f^2a^4} \left\{\begin{array}{ll} \frac{4|c_1|^2}{(\ga+1/2)^2}x^{2\ga+2} & \mbox{if } \!\ga
   \!<-1/2 ,\\
\hspace{-0.2cm}(1\!-\!2\ga)^2|c_2|^2x^{-2\ga} & \mbox{if } \ga\! >\!-1/2 \end{array} \right. \label{PE}\\
\hspace{-0.2cm}P_{EB} (k,\eta) &=& \frac{4\pi k}{f^2a^4}   \left\{\begin{array}{ll} \frac{-2|c_1|^2}{\ga+1/2}x^{2\ga+1} & \mbox{if } \ga < \!-1/2 \\
\hspace{-0.2cm} (2\ga-1)c^*_1c_2& \hspace{-0.2cm} 
\mbox{if } -1/2\!<\ga\!<\!1/2 \nonumber\\
\hspace{-0.2cm} (2\ga-1)|c_2|^2x^{1-2\ga} &\mbox{if } \ga\!>\!1/2\,. \end{array} \right.\\
\label{PEandB}
\eea  
As an example, we compute the spectrum of the anisotropic stress generated by the magnetic field $P_B$, i.e. the second term in the sum \eqref{ppiP2}. For $\gamma<1/2$, we take the first line in Eq.~\eqref{PB}. Using the general formula Eq.~\eqref{e:Iapprox} from
Appendix~\ref{app:conv} to approximate the convolution, we obtain
\bea
\label{e:PB1}
P_{\Pi^{(B)} } & \simeq& \frac{9|c_1|^4 \si_2}{4\pi^2|\eta|^5a^8}  \\
& \times & \left\{\begin{array}{ll}
\frac{2\ga+1}{(4+2\ga)(5+4\ga)}\, x^{5+4\ga} & \mbox{if } -2<\ga<-{5}/{4} \,,\\    
\frac{1}{5+4\ga}  & \mbox{if } -5/4<\ga<1/2 \,. \nonumber
\end{array}\right.
\eea
For $\gamma>1/2$, we take the second line in Eq.~\eqref{PB} and we obtain
\be\label{PB2}
P_{\Pi^{(B)}} \simeq\frac{9|c_2|^4 \si_2}{4\pi^2|\eta|^5a^8}\,\frac{1}{9-4\ga}\,, \qquad \mbox{if } \ga >1/2 \, .
\ee
Similar expressions for the anisotropic stress generated by the electric field
 and the cross term are computed in Appendix~\ref{app:PEandB}. 

From these expressions, we can evaluate $P_\Pi(k,\eta)$. 
Comparing the scaling of the magnetic, electric and cross term contributions, we can identify three different regimes.
For $-2<\gamma<-5/4$, the magnetic field always dominates for $x<1$, and we can neglect the electric contribution
and the cross term.
For $-5/4<\ga<5/4$ all the contributions are of the same order of magnitude. This follows from the fact that for these values of $\ga$
the integrals over $k'$ are dominated by the upper bound $1/|\eta|$, leading to a white noise spectrum (see Appendix~\ref{app:conv}).
Finally, for $5/4<\ga<2$, the electric field contribution dominates and in principle we could neglect the magnetic field contribution 
and the cross term. However, at the end of inflation, when the Universe enters the radiation 
era, conductivity quickly becomes very high,  meaning that the electric field decays rapidly
 (see for example~\cite{AE}). The only remaining contribution to the 
anisotropic stress then is due to the magnetic field. Therefore, in this case we  
keep both the electric and the magnetic contribution in the anisotropic stress.
Putting everything together we find the following power spectrum for the scalar anisotropic 
stress potential:

\begin{widetext}
\be
\label{Ptot}
P_{\Pi}(k,\eta) \simeq\frac{9}{4\pi^2|\eta|^{5}a^{8}}\left\{\begin{array}{ll}
\frac{|c_1|^4\si_2(1+2\ga)}{(4+2\ga)(5+4\ga)}\,x^{5+4\ga} &  \mbox{ if } -2<\ga<-5/4\\    \\
\frac{|c_1|^4\si_2}{5+4\ga}  +\frac{|2c_1|^4\si_1}{(1/2+\ga)^4(9+4\ga)} +\frac{4|c_1|^4\si_3}{(1/2+\ga)^2(7+4\ga)}  
& \mbox{ if } -5/4<\ga<-1/2 \\  \\
\frac{|c_1|^4\si_2}{5+4\ga}  +\frac{|(1-2\ga)c_2|^4\si_1}{5-4\ga}+\frac{\si_3|c_1c_2|^2(1-2\ga)^2}{5} &  \mbox{ if } -1/2<\ga<1/2 \\  \\
\frac{|c_2|^4\si_2}{9-4\ga}+\frac{|c_2|^4\si_1(1-2\ga)^4}{5-4\ga}+\frac{|c_2|^4 \si_3(1-2\ga)^2}{7-4\ga}   &
\mbox{ if } ~ 1/2<\ga<5/4 \\  \\
\frac{|c_2|^4\si_1(1-2\ga)^5}{(4-2\ga)(5-4\ga)}\, x^{5-4\ga} + \frac{|c_2|^4\si_2}{9-4\ga} &
\mbox{ if } ~ 5/4<\ga<2 \,.
\end{array}\right.
\ee
\end{widetext}
The expressions above diverge at the boundary of their validity. Except in the case $|\ga|=2$, this is simply because 
our approximation for the integrals derived in Appendix~\ref{app:conv} breaks down: the true integral would remain finite. For $|\ga|=2$ instead, this is the usual log divergence of a scale-invariant spectrum. 
Note that the magnetic field contribution for $\ga>5/4$ (the term proportional to $\si_2$ in the last line of the above equation) is smaller than the one from the electric field by a factor $x^{4\ga-5}$, for $x<1$. Hence the effect of this contribution to the dynamical evolution of the Bardeen potential during inflation is negligible with respect to the effect of the electric field: we can therefore neglect it in the computation of the source during inflation. However, as previously mentioned, only the magnetic field survives beyond inflation and generates the anisotropic stress, because of the high conductivity during the radiation era. We have therefore to take into account the magnetic field contribution for the solution of the Bardeen equation in the radiation era, see Section \ref{Bradera}.  

Furthermore, note that the prefactor $|\eta|^{-5}$ has the correct dimension: since $[\pis(\bx)] = [\ell^{-4}]$ and the Fourier transform is a volume integral, we have $$[\ell^{-2}] =[\langle\Pi_S^\dagger(\bk)\Pi_S(\bk')\rangle] =
[\de(\bk-\bk')P_\Pi]\,.$$
With $ [\de(\bk-\bk')] = [\ell^{3}]$ this implies $[P_\Pi] =[\ell^{-5}]$.

For the remainder of this paper, we use a simplified expression for the power spectrum $P_\Pi$. For all values of $\gamma \in [-2,2]$, Eq.~\eqref{Ptot}  is of the following form, for superhorizon scales $x<1$:
\be\label{e:PPisgen}
P_\Pi(k,\eta) \simeq \frac{C^2_\Pi(\ga)}{a^8|\eta|^5}\left\{\begin{array}{ll} 
x^{5-4|\ga|} & \mbox{if } 5/4<|\ga| <2 \\
1 &\mbox{else,}
\end{array}\right.
\ee
where $C_\Pi(\ga)$ is a dimensionless parameter of order unity. This expression for $P_\Pi$ allows us to simplify considerably the analytical evaluation of the metric perturbations. It involves several approximations, but the spectral shape is correct and the evaluation of the prefactor is the best possible if one proceeds analytically. 

In order to compute the source of Eq. \eqref{sourceLS}, we also need to determine 
$\Pi_S'(\bk,\eta)$; expression (\ref{e:PPisgen}) helps us in this task. 
We note in fact that on superhorizon scales and at lowest order in the slow-roll expansion, the time 
dependence of the power spectrum of $\Pi_S(\bk,\eta)$ is given by $P_\Pi \propto |\eta|^{2\al}$ with
\bea\label{e:alpha}
\al(\ga) &=& \left\{\begin{array}{ll}  
               4-2|\ga| \, , & |\ga| \ge 5/4 \\
               3/2 \, , &  |\ga|\le 5/4 \end{array}\right.\\
            &=& \min\{4-2|\ga|,3/2\} \,.
\eea
Moreover, from Eqs.~\eqref{pis} and \eqref{Aapp} one finds that on large scales $x<1$, the operator $\Pi_S(\bk,\eta)$, expressed in terms of the variables $x$ and $\bk$, is simply a power law in $x$. Recalling Eq.~\eqref{e:SOp}, we can simply write 
$\Pi_S(\bk, x) = x^m \sum_{i=1}^4{\bar \Pi}_i(\bk)O_i(\bk) $, where the entire time dependence is collected in the prefactor. Hence the power spectrum must go like $x^{2m}$: 
\be
(2\pi)^3\de(\bq-\bk)P_\Pi(k) = x^{2m}|{\bar \Pi}_1(\bk)|^2\langle 0|O_1^\dagger(\bq)O_1(\bk)|0\rangle \,,
\ee
setting $i=1$ for the only kind of operator which survives. 
Eq.~\eqref{e:PPisgen} therefore implies $m=\al$, so that we find the simple relation valid at lowest order in the slow-roll expansion: 
\be
\Pi'_S(\bk,\eta) \simeq - \frac{\al}{|\eta|} \, \Pi_S(\bk,\eta)\,, \qquad {\rm at}~\mathcal{O}(\epsilon^0)\,.
\ee
Note that at next order in slow-roll, $\Pi_S(\bk,\eta)$ gets an extra time dependence as $|\eta|^{4\epsilon}$. We now have all the ingredients to determine the source term in Eq.~(\ref{sourceLS}) at lowest order in slow-roll: 
\be
\SB\simeq \frac{3}{x^4} \big(2-\alpha\big) \frac{\Pi_S(\bk,\eta)}{\rho_\varphi}\,.
\label{sourceLSep}
\ee
In Section \ref{sec:curvature}, to integrate the $\zeta$-equation \eqref{eq:zeta}, we shall also need the spectrum of the electromagnetic energy density and its derivative. These can be obtained along  the same lines  as above, with the same qualitative results. We are not repeating the details. Defining as usual
\be
\langle 0| \rho_\eb^\dagger(\bq,\eta) \rho_\eb(\bk,\eta)|0\rangle = (2\pi)^3P_{\rm em}(k,\eta)\de(\bq-\bk)\,,
\ee
and substituting Eqs.~\eqref{expansion} and~\eqref{Asolution} in $\rho_{\eb}(\bx,\eta)=-T^0_{\eb\,\,0}(\bx, \eta)$, one obtains an expression analogous to Eq.~\eqref{ppiP2}, where only the constants $\si_1(\ga),~\si_2(\ga)$~and~$\si_3(\ga)$ are different. The final result can again be parametrized as
\be\label{e:Pemgen}
P_{\rm em}(k,\eta) \simeq \frac{C^2_{\rm em}(\ga)}{a^8|\eta|^5}\left\{\begin{array}{ll} 
x^{5-4|\ga|} & \mbox{if } 5/4<|\ga| <2 \\
1 &\mbox{else,}
\end{array}\right.
\ee
leading to 
\be\label{e:rhoem'}
\rho'_\eb(\bk,\eta) \simeq - \frac{\al}{|\eta|} \, \rho_\eb(\bk,\eta)\,, \qquad {\rm at}~\mathcal{O}(\epsilon^0)
\ee
with $\al$ as defined in Eq.~\eqref{e:alpha}.

With this, we can now proceed to solve Bardeen equation. 

\section{Resolution and matching}
\label{sec:solution}

\subsection{Bardeen potentials during inflation}
\label{sec:bardeeninflation}

We now want to compute $\Psi$ at large scale by solving Eq.~\eqref{psiroll} with the source term given by 
Eq.~\eqref{sourceLSep}. 
The homogeneous solutions of Eq.~\eqref{psiroll} are $\Psi_1=x^pJ_\nu(x)$ and $\Psi_2=x^pJ_{-\nu}(x)$, with $p=1/2+\mathcal{O}(\ep,\ep_2)$ and
$\nu=1/2+\mathcal{O}(\ep,\ep_2)$. 
The solution of the inhomogeneous equation can be computed using the Wronskian method, and reads 
\be
\begin{split}
\label{psiinh}
\hat \Psi_{\rm inh}(\bk, x) = & \frac{-\pi}{2\sin(\nu\pi)} \int_{x_{\rm in}}^x  dx' x'
S_{\rm em}(\bk, x')\left(\frac{x}{x'}\right)^p \\
& \bigg[J_\nu(x')J_{-\nu}(x)-J_\nu(x)J_{-\nu}(x') \bigg]\; ,
\end{split}
\ee
where $x_{\rm in}=|k\eta_{\rm in}|$ is the initial time when the source starts to act, i.e. horizon exit 
$x_{\rm in}\sim 1$, and it is such that $x_{\rm in}\gg x$. 
We denote the Bardeen potential by $\hat \Psi_{\rm inh}$ in order to indicate that this expression is a quantum operator
acting on the electromagnetic vacuum: we will then have to relate it to a stochastic variable after inflation in the usual 
way, see the discussion later in this section.  Also in this case, the solution becomes highly 
squeezed after horizon exit. Only the dominant mode of the Bessel function,  $\Psi_2=x^pJ_{-\nu}(x)$ remains relevant.
Since we are interested in the large-scale solution $x\ll x_{\rm in}\lesssim 1$, we can expand the Bessel functions in Eq.~\eqref{psiinh} for small arguments. Moreover, to integrate Eq.~(\ref{psiinh}) we only need to know the time 
behavior of the source term, which is proportional to $ x^{\al-4}$ (c.f. Eq.~\eqref{sourceLSep}). We obtain 
\bea    \label{e:Psinh}
\hat\Psi_{\rm inh}(\bk, x) &\simeq& \frac{x^2}{(2-\al)(3-\al)}S_{\rm em}(\bk,x)  \\
   &=& \frac{ \beta}{ x^2}\frac{\pis}{\rho_\varphi} \,.
\eea
Note that, since $0\le \al\le 3/2$, the prefactor 
\be
 \beta = \frac{3}{(3-\al)} 
\label{beta}
\ee
is positive for all values of $\ga$. 
Neglecting the decaying mode of the homogeneous solution $\Psi_1\simeq \sqrt{2\pi} \,x/4$, we obtain 
the general large-scale solution for the Bardeen potentials during inflation,
\bea
\label{psiminusO}
\hat\Psi_{-} &\simeq& \hat b(k) + \frac{\beta}{x^2}\frac{\Pi_S}{\rho_\varphi}\quad \mbox{and}\\
 \hat\Phi_{-} &\simeq& \hat b(k) + \frac{\beta-3}{x^2}\frac{\Pi_S}{\rho_\varphi}\; , \label{phiminusO}
\eea
where $\hat b(k)$ is the usual inflationary solution at large scales, the homogeneous `growing mode' (which is constant in time). To 
obtain Eq.~\eqref{phiminusO} we have used Eq.~\eqref{phiminuspsi} which yields
\be
\hat\Phi_{-} = \hat\Psi_- -\frac{3}{x^2}\frac{\Pi_S}{\rho_\varphi} \,.
\ee  

The above solutions are the sum of two uncorrelated 
quantum operators. The first, $\hat b$, acting on the inflaton vacuum, and the second, proportional to $ \Pi_S$, acting on the electromagnetic vacuum. As mentioned before, these quantum variables must be identified with classical perturbations having stochastic amplitudes by means of a quantum to classical transition: this is explained, for example, in \cite{gr-qc/9504030,gr-qc/9806066}. For simplicity we identify these variables with their r.m.s. amplitude, i.e. the square root of the volume factor in wave number space times their corresponding spectra: this is not a conventional choice but it allows to simplify considerably our formulas. From the corresponding power spectra, defined as 
\bea 
\hspace{-0.5cm} \langle 0| \hat b^\dagger(\bq,\eta) \hat b(\bk,\eta)|0\rangle &=& (2\pi)^3P_b(k,\eta)\de(\bq-\bk)\,, \\
\hspace{-0.5cm} \langle 0|\hat\Psi_-^\dagger(\bq,\eta)\hat\Psi_-(\bk,\eta)|0\rangle &=& (2\pi)^3P_\Psi (k,\eta)\de(\bq-\bk)\,,
\eea
we therefore define the dimensionless metric perturbations:
\bea 
\hat\Psi_- \rightarrow \Psi_-=\sqrt{k^3P_\Psi} \nonumber \\
\hat b \rightarrow b=\sqrt{k^3P_b}\,. \nonumber
\eea
An advantage of this somewhat unconventional definition is that both $\Psi_-(k,\eta)$ and $b(k,\eta)$
are dimensionless and provide a good measure for the fluctuation amplitude at comoving scale $k$.
Analogously, for the anisotropic stress power spectrum we introduce the dimensionless ratio $\hpis$ by
\bea
\left[\hpis(k,\eta)\right]^2 &\equiv& \frac{k^3 P_\Pi}{\rho^2_\vph} = \left(\frac{8\pi G}{3H^2}\right)^2 k^3P_\Pi \\
\hpis(k,\eta) &\simeq & \frac{H^2}{3 m_P^2}C_\Pi(\ga) x^\al \label{hpis}\, .
\eea
The superscript ${}^-$ indicates that we evaluate the quantity in the inflationary era (as opposed to the radiation era, see Section \ref{Bradera}). Note that here the Hubble parameter is taken at lowest order in slow-roll: $H=\HH/a\simeq 1/(a_1\eta_1)$, c.f. Eqs.~\eqref{aHdef}. 
With this definition, the power spectrum of the source term becomes
\bea
&& \hspace{-0.5cm} \langle 0|S_{\rm em}^\dagger(\bq,\eta)S_{\rm em}(\bk,\eta)|0\rangle = (2\pi)^3P_S(k,\eta)\de(\bq-\bk)\,, \\
&& k^3 P_S(k,\eta) = 9(2-\al)^2\left(\frac{\hpis}{x^4}\right)^2\; .
\eea
With Eqs.~\eqref{e:Psinh} and \eqref{beta}, we find the relation among the power spectra 
\be   \label{e:PsinhS}
k^3 P_{\Psi_{\rm inh}} \simeq \frac{x^4}{(2-\al)^2(3-\al)^2} k^3P_S = \beta^2 \left(\frac{\hpis}{x^2}\right)^2\,. 
\ee
We can now rewrite the solutions Eqs.~\eqref{psiminusO} and \eqref{phiminusO} in terms of the classical variables, which we understand as their r.m.s. amplitudes, both for the inhomogeneous part of the solution and for the inflationary part of the solution. We obtain  with Eq.~\eqref{e:PsinhS}
\bea
\label{psiminus}
\Psi_{-}(x) &\simeq& b(k) + \beta \, \frac{\hpis}{x^2}\quad \mbox{and}\\
 \Phi_{-}(x) &\simeq& b(k) + (\beta-3)\, \frac{\hpis}{x^2}\; , \label{phiminus}
\eea
within the approximation that the magnetic field perturbations and the inflaton perturbations are uncorrelated,
$\langle 0| \Psi_{\rm inh}^\dagger(k) \hat b(k)|0\rangle =0$ (note that this is violated in second-order perturbation 
theory~\cite{1109.4415}). Therefore, the power spectra of $\Psi_-$ and $\Phi_-$ are simply the sum of the 
inflationary power spectrum and the inhomogeneous one.

The inhomogeneous part of the solutions Eqs.~\eqref{psiminus} and \eqref{phiminus}, $ \hpis/x^2 $, 
behaves like $x^{\al(\ga)-2}$. If $\ga=-2$, i.e. when the magnetic field energy density generated during inflation has a scale-invariant spectrum, one has 
$\al=0$ and therefore the inhomogeneous mode grows in time like $x^{-2}$.
One may wonder whether this leads to too large metric fluctuations, but it is not the case: considering  
the ratio of the Weyl tensor 
${C^\mu}_{\nu\al\beta}\propto k^2(\Phi+\Psi) $ and the Ricci tensor, 
$R_{\mu\nu} \propto \HH^2 \simeq \eta^{-2}$ one finds for the ratio of typical components of the Weyl,
respectively Ricci, tensor~\cite{mybook}
\be
\left|\frac{\rm Weyl}{\rm Ricci}\right|_{\rm inh} \simeq x^2(\Phi_{\rm inh}+\Psi_{\rm inh})ì \sim \hpis \ll 1 \,.
\label{WeyloverRicci}
\ee
The last inequality is a consequence of the fact that we require the electromagnetic field to be subdominant during the inflationary era, so that $\pis\simeq \rho_{\rm em}\ll\rho_{\vph}$.

\subsection{Bardeen potentials during the radiation era}
\label{Bradera}

After inflation and reheating, the Universe enters the radiation-dominated phase. Filled by a fully ionized 
plasma of relativistic particles, the Universe becomes conductive, in contrast to the inflationary phase during which there are no free charges. The conductivity of the Universe is very high, so that the electric field disappears  rapidly~\cite{yokoyama,subramanian,AE}. Wavelengths of cosmological interest are  much larger than the horizon 
scale at the end of inflation. For them, the transition to the radiation era and the dissipation of the electric field can be considered as instantaneous. 

The evolution of the Bardeen potential in the radiation era has already been studied in detail in~\cite{cc} and \cite{shaw}. Einstein's equations can again be combined into a second-order equation for 
$\Psi$, that reads (see Eq.~(B5) in~\cite{cc})
\be
\label{psirad}
\Psi''+4\HH\Psi'=\frac{3\HH}{k^{2}}\big(\HH^2\hpisp\big)'\; ,
\ee
where again we identify the metric perturbations with their r.m.s. amplitudes, and 
$\hpisp=\sqrt{k^{3}P_\Pi}/{\bar\rho_{\rm rad}}$ is the dimensionless magnetic anisotropic stress parameter in the radiation era. It is constant in time 
during the radiation era as both, the radiation density and $B^2$ scale as $a^{-4}$.
Its value depends on $\ga$ and is given by the magnetic field contribution of Eq.~\eqref{Ptot}, i.e. by the part proportional to
$\si_2$, evaluated at $\eta=\eta_*$. This magnetic part is indeed the only one that survives in the highly conductive  radiation era.
The general solution to Eq.~\eqref{psirad} at large scales is 
\bea
\Psi_{+}(x) &=& \Psi_0+\frac{\Psi_1}{x^3}+\frac{3\hpisp}{x^2}\quad \mbox{and}  \nonumber\\
\Phi_{+}(x) &=& \Psi_0+\frac{\Psi_1}{x^3} \; ,
\label{psiplus}
\eea
where $\Psi_0$ and $\Psi_1$ are two arbitrary constants that need to be determined by matching the 
solutions in the radiation era to the one during inflation.

\subsection{Matching}
\label{sec:matching}

We have found the solutions for the metric potentials $\Psi$ and $\Phi$ in the presence of an electromagnetic field both during inflation (at lowest order in the slow-roll expansion) and in the radiation era. The solutions in the radiation era are known once the initial conditions are specified. In order to have solutions valid through the whole evolution of the Universe, we need to match properly $\Psi$ and $\Phi$ at the transition from inflation to the radiation era. 

The initial conditions in the radiation era are obtained by matching the solutions given in
Eqs.~\eqref{psiminus} and \eqref{phiminus} to Eqs.~\eqref{psiplus}. As usual, we match the classical potentials among themselves. 
We are only interested in wavelengths much larger that the duration of the transition. For these scales, the transition can be considered as instantaneous. Within this approximation, the equation of state 
 parameter $w$ experiences a discontinuity at the transition, it goes from roughly $w\simeq-1$ to $w=1/3$. 
Furthermore, the electromagnetic field anisotropic stress is discontinuous when $\gamma>-5/4$, due to the fact that the electric field contributes significantly during inflation for these values of $\gamma$, while it vanishes in
the highly conductive plasma of the radiation era. 

In \cite{muk_deruelle} it has been shown that to match solutions through a discontinuity of the energy-momentum tensor, we have to impose that  the induced 3-metric and the extrinsic curvature remain continuous on the spacelike hypersurface of the transition $\Sigma$. We follow this procedure here.

The most general metric containing only scalar perturbations is given by
\be
\begin{split}
ds^2=&a^2\bigg\{-(1+2A)d\eta^2+2B_{,i}d\eta dx^i \\
&+\big[(1+2C)\delta_{ij}+2E_{,ij} \big]dx^idx^j\bigg\}\; .
\end{split}
\ee
A convenient choice of coordinates to fix the matching conditions is to define the transition hypersurface 
through $\tilde{\eta}={\rm const}$.
The time coordinate $\tilde\eta$ is related to the original one by a gauge transformation 
$\tilde\eta=\eta+T$~\cite{muk_deruelle}. 
On the $\{\tilde\eta = {\rm const.}\}$  slices, the continuity of the induced 3-metric and of the extrinsic curvature requires the continuity of $\tilde E$, $\tilde C$, $\tilde B-\tilde E'$ and $\tilde C'-\HH \tilde A$ through the transition~\cite{muk_deruelle}.
Considering the gauge transformation properties of the different metric components~\cite{mybook},
this implies in terms of the original perturbation variables
\begin{align}
\label{matchgen}
&[E]_\pm=0\;,&  &[C+\HH T]_\pm=0\;,\\
& [B-T-E']_\pm=0\;, &  &[-\HH A+C'-(\HH^2-\HH')T]_\pm=0\;,\nonumber
\end{align}
where 
\be
\quad F_\pm=\lim_{\varepsilon\rightarrow 0}[F(+\eta_*+\varepsilon)-F(-\eta_*-\varepsilon)]\; ,
\ee
$\eta_*$ being the time of the transition: the end of inflation happens at $\eta=-\eta_*$ and the radiation phase is established at $\eta=+\eta_*$. This also ensures the continuity of the Hubble parameter, the 
unperturbed extrinsic curvature: at lowest order in the slow-roll expansion one has 
 $$\HH_{\rm inf}(\eta=-\eta_*) = \eta_*^{-1} = \HH_{\rm rad}(\eta=\eta_*)\equiv \HH_*\,.$$ 
 Our original gauge is longitudinal gauge with $E=B=0$, $A=\Phi$ and $C=-\Psi$. With this 
 Eqs.~\eqref{matchgen} become
\begin{align}
\label{matchlong}
&[T]_\pm=0\;, \quad [\Psi]_\pm=0\;,\\
&[\HH \Phi+\Psi'+(\HH^2-\HH')T]_\pm=0\;.\nonumber
\end{align}
Inserting the solutions Eqs.~\eqref{psiminus} and \eqref{psiplus} into Eq.~\eqref{matchlong}, and using that $(\HH^2-\HH')_{-}=\ep\HH_*^2$ 
and $(\HH^2-\HH')_{+}=2\HH_*^2$, one can determine the constants $\Psi_0$ and $\Psi_1$ in terms of $T$. Note that we set $(\HH^2-\HH')_{-}=\ep\HH_*^2$ even though we are performing the matching at lowest order in slow-roll, since the order of $T$ is as far not determined.
A brief computation yields
\bea
\Psi_{0} &= & b(k)+\frac{\ep-2}{3}\HH_*T+\frac{2\beta}{3}\frac{\hpis(x_*)}{x_*^2} 
  \; ,\label{Psi0T} \\
\frac{\Psi_{1}}{x_*^3}&=& -\frac{\ep-2}{3}\HH_*T + \frac{\beta}{3}
   \frac{\hpis(x_*)}{x_*^2}  - 3\frac{\hpisp(x_*)}{x_*^2} \;.
\eea

If one inserts the above constants in the radiation solution Eqs.~\eqref{psiplus},  the following problem for the metric perturbations becomes manifest: since $0<\al<3/2$, the term 
$$\frac{\hpis (x_*)}{x_*^2} \sim  \left(\frac{H_*}{m_P}\right)^2x_*^{\al-2} \geq  \left(\frac{H_*}{m_P}\right)^2x_*^{-1/2}\; , $$ 
which can be very large for small $x_*$,
enters in the constant mode $\Psi_0$ of both $\Psi_+$ and $\Phi_+$, and contributes to the fluctuation amplitude as
$$\sqrt{k^3 P_{\Psi_0}} \sim\frac{\hpis(x_*) }{x_*^2}~.$$
If this term is not `compensated' in any way, it leads to very large fluctuations. Indeed, if one evaluates again the same ratio as in Eq.~\eqref{WeyloverRicci} but this time in the radiation era, one finds
\be
\left|\frac{\rm Weyl}{\rm Ricci}\right|_{\rm rad} \sim \hpis(x_*)\left(\frac{x}{x_*}\right)^2 \; ,
\label{WoverRrad}
\ee
which can become very large with the expansion of the Universe. However, we now show that this term is exactly compensated by the term proportional to $T$ in Eq.~(\ref{Psi0T}), if we choose a proper hypersurphace for the transition and deal with the slow-roll expansion properly.

To go on we need to specify a physically meaningful transition hypersurphace $\Sigma$, in order to determine $T$. 
Since inflation is driven by the scalar field, which controls the background evolution of the Universe, it is reasonable to assume that inflation ends when the energy density of the scalar field reaches a certain value. 
In this case, the hypersurface of transition is the one of constant energy density of the scalar field, such that $\rho_{\varphi}(\tilde \eta,\bx)=\bar{\rho}_{\varphi}+\delta\rho_{\varphi}={\rm const}$, at constant $\tilde \eta$. This requires to choose $T=-\delta\rho_{\varphi}/{\bar{\rho}_{\varphi}'}$, so that $\widetilde{\delta\rho_{\varphi}}=0$ \cite{muk_deruelle}. We can compute $T$ explicitly at the end of inflation, for $\eta=-\eta_*$. Using the scalar Einstein equations to 
eliminate the inflaton perturbations we find
\be\label{Trhofix}
\HH_*T=\frac{-1}{3(1+w_-)}\bigg[2\Phi_-+\frac{2\Psi'_-}{\HH_*}+\Omega_\eb^-+\frac{2k^2\Psi_-}{3\HH_*^2} \bigg] \; .
\ee
Here $\Omega_\eb^-=\rho_{\rm em}/\bar\rho_\varphi$ is the density parameter of the electromagnetic field which is of
the order of $x_*^2\Psi_-$ like the last term in Eq.~\eqref{Trhofix}: these two terms can therefore be neglected at lowest order in $x_*\ll 1$. 
The first two terms in Eq.~\eqref{Trhofix} are related to the curvature perturbation $\zeta$ 
defined in Eq.~(\ref{zeta}). Since $\Psi$ and $\HH$ do not jump, to lowest order in $x_*$
the jump in $\zeta$ is related to the one in $T$ by $\HH_*[T]_\pm = -\left[\zeta  \right]_\pm$. The condition 
$[T]_\pm = 0$ for the $\{\rho_\varphi = {\rm const.}\}$ hypersurface then implies
\be
\left[\zeta  \right]_\pm =0 \,.
\label{zetapm}
\ee
Hence the constant density matching requires that the curvature perturbation is continuous at lowest order in $x_*\ll 1$ (but it can be discontinuous at next-to-leading order). Inserting 
Eqs.~\eqref{psiminus} and ~\eqref{phiminus} into Eq.~\eqref{Trhofix}, we find to lowest 
order in $x_*\ll 1$ and in the slow-roll expansion
\be
\label{Tleading}
\HH_*T\simeq -\frac{b(k)}{\ep}\;.
\ee
It appears therefore that, if the hypersurface of transition is the one of constant energy density for the inflaton, $T$ is of $\mathcal{O}(\epsilon^{-1})$ at lowest order in the slow-roll expansion. With this choice for $T$, the solution Eq.~\eqref{psiplus} in the radiation era for $x_*\ll 1$ is completely determined, with coefficients: 
\bea
\Psi_{0}&=& \frac{2}{3}\left(1+\frac{1}{\ep}\right)b(k)
 + \frac{2\beta}{3}\frac{\hpis }{x_*^2} \,, \label{Psizero} \\ 
\frac{\Psi_{1}}{x_*^3}&=&\frac{1}{3}\left(1-\frac{2}{\ep}\right)b(k)  + \frac{\beta}{3}\frac{\hpis }{x_*^2} - \frac{3\hpisp}{x_*^2} \label{Psi1}\;.
\eea
From these equations we see that the magnetic field anisotropic stress does not contribute to the leading order in slow-roll $\mathcal{O}(\epsilon^{-1})$ . 
The following order $\mathcal{O}(\epsilon^{0})$ cannot be trusted: this order would indeed receive contributions from $T$ at next-to-leading order in slow-roll that we did not take into account in Eq.~\eqref{Tleading}.
To be consistent it is therefore necessary to solve Bardeen equation up to next-to-leading order in the slow-roll expansion, as anticipated in Section \ref{sec:metric}. 

To solve Eq.~(\ref{psiroll}) at next-to-leading order in the slow-roll expansion, we proceed as in Section \ref{sec:bardeeninflation}. In the following we present the solution and the matching for the inhomogeneous part only, sourced by the electromagnetic field; the relevant inflationary contribution in the radiation era is the one given in  Eqs.~(\ref{Psizero}), (\ref{Psi1}) at lowest order in slow-roll. For the inhomogeneous solution at next-to-leading order, we have to account for the fact that the source Eq.~(\ref{sourceLSep}) has now a time dependence of the form $x^{\alpha+2\epsilon}$. We find the inhomogeneous part of the solution 
\be
\hat\Psi_{{\rm inh}}^- \simeq ( 1+ 3 \epsilon) \frac{\beta}{x^2} \frac{\Pi_S}{\rho_\varphi}\,.
\ee
We then identify the Bardeen potential quantum operator with its r.m.s. amplitude as previously done, and we find the expressions (equivalent to Eqs.~\eqref{psiminus}, \eqref{phiminus} up to next-to-leading order $\mathcal{O}(\epsilon)$):
\bea
&\Psi_{\rm inh}^-(x) \simeq  \beta \left[1+ \Big(1+2\log \big(\frac{\eta}{\eta_1} \big) \Big) \epsilon \right] \frac{\hpis}{x^2}& \label{psiminusep} \\
&\Phi_{\rm inh}^-(x) \simeq  \left[ \beta-3 + \Big(\beta+2(\beta-3) \log(\frac{\eta}{\eta_1}\big) \Big) \epsilon \right] \frac{\hpis}{x^2} &\label{phiminusep}
\eea
The logarithmic factors arise because we have kept definition \eqref{hpis} for $\hpis$, which is at lowest order in slow-roll. 

The matching proceeds as before with the difference that the continuity of the unperturbed extrinsic curvature $\HH$ implies that the radiation phase is now established at time $\eta=\eta_*/(1+\epsilon)$: 
\be
\HH_{\rm inf}(\eta=-\eta_*) = \frac{1+\epsilon}{ \eta_*} = \HH_{\rm rad}\big( \eta=\frac{\eta_*}{ 1+\epsilon}\big) \equiv \HH_*\,.
\ee
Inserting solutions (\ref{psiminusep}) and (\ref{phiminusep}) into Eq.~\eqref{Trhofix}, we obtain for the electromagnetic contribution to the gauge transformation variable $T$ at next-to-leading order:
\be
\label{Teps}
\HH_*T\simeq -\frac{b(k)}{\epsilon}+  \beta \, \frac{\hpis }{x_*^2} \;.
\ee
Inserting this expression in Eq.~\eqref{Psi0T}, we find that at next-to-leading order, $\HH_*T$ cancels exactly the magnetic mode in $\Psi_0$ given by $(2\beta/3)\,\hpis(x_*)/x_*^2$ (c.f. Eq.~\eqref{Psizero}), so that in $\Psi_0$ only the inflationary contribution remains. {\it The matching conditions, Eqs.~\eqref{matchlong}, insure therefore that the magnetic mode $\propto \hpis / x^2$ does not transfer into the constant mode $\Psi_0$ at the transition to the radiation era.}~Note that, in order to reach this conclusion, it is absolutely necessary to perform the matching. The decaying mode of the solution for $\Psi_+$ in the radiation era which is proportional to $\Om_\Pi^+$ does not give any information on $\Psi_0$ (c.f. Eq.~\eqref{psiplus}). Therefore, the calculation done in Ref.~\cite{suyama} and, in particular, their solution Eq.~(33) are not sufficient to claim that no electromagnetic contamination to $\Psi_0$ is present after inflation.  

As a result of the matching, we see that the dangerous behavior inferred in \eqref{WoverRrad} is not present. Since the decaying mode in Eqs.~\eqref{psiplus} can be neglected, we arrive at the result that the inflationary electromagnetic field contributes to the metric perturbations in the radiation era through $\Psi_0$ only at next-to-leading order in $x\ll 1$, i.e. at order $\mathcal{O}(x^0)$ instead of $\mathcal{O}(x^{-2})$. Therefore, in order to find the relevant effect of the inflationary electromagnetic field, we would have to go to the next-to-leading order in the large-scale expansion. However, solving Bardeen Eq. (\ref{psiroll}) analytically at next-to-leading order in $x=k|\eta|\ll 1$ and in the slow-roll expansion is highly nontrivial. The electromagnetic contribution to the metric perturbations can be found more easily by solving for the curvature perturbation $\zeta$ which, as shown at the end of Section \ref{sec:metric}, is only sourced at next-to-leading order in $x\ll 1$.  

\subsection{The curvature perturbation}
\label{sec:curvature}

The metric perturbations in the radiation era are given in Eqs.~\eqref{psiplus}: neglecting the decaying mode, the relevant contribution generated by the electromagnetic field is in the constant mode $\Psi_0$, but only at order $\mathcal{O}(x^0)$ as demonstrated in the previous section. The simplest way to obtain $\Psi_0$ is to solve for the curvature perturbation $\zeta$. Indeed, inserting Eqs.~\eqref{psiplus} into the definition of $\zeta$~Eq.~\eqref{zeta} one finds that
\be
\zeta_{+}=\frac{3\Psi_0}{2}\; .
\ee   
Since $\zeta$ is continuous at the transition from inflation to the radiation era, as demonstrated in Eq.~\eqref{zetapm}, this means that $\Psi_0$ is given by the value of the curvature at the end of inflation
\be
\Psi_0=\frac{2\zeta_-}{3}\equiv \frac{2}{3}\left[\left( 1 +\frac{1}{\ep}\right)b(k) +\zeta_*\right]\; ,
\ee
where we denote by $\zeta_*$ the contribution sourced by the electromagnetic field during inflation, which we need to determine.  

Note that the solutions Eqs.~\eqref{psiminus} and~\eqref{phiminus}  do not allow to compute $\zeta_*$.   
Indeed, inserting them into Eq.~\eqref{zeta} we find that the term $\HH\Phi+\Psi'$ vanishes. Naively one might think that this implies simply $\zeta_-=\Psi_-$, and that $\zeta$ has a contribution at order $\mathcal{O}(x^{-2})\cdot \mathcal{O}(\epsilon^{0})$; however, this would be in contradiction with Eq.~\eqref{zetafirst}, which shows that $\zeta$ is only sourced at next-to-leading order in $x\ll 1$. In reality, this contradiction is solved if one inserts in definition~\eqref{zeta} the Bardeen potentials at order $\ep$ and not simply $\ep^0$, because the term $\HH\Phi+\Psi'$ enters with a prefactor of order ${\cal O}(1/(w+1)) ={\cal O}(\ep^{-1})$ in \eqref{zeta}. Inserting solutions ~\eqref{psiminusep} and~\eqref{phiminusep} one verifies that the order $\mathcal{O}(x^{-2})\cdot \mathcal{O}(\epsilon^{0})$ vanishes, and we expect that the same thing happens at any following order in $\epsilon$ (note that Eq.~\eqref{zetafirst} is indeed valid at any order in slow-roll). Consequently, the relevant contribution of the electromagnetic field $\zeta_*$ is of the order $\mathcal{O}(x^{0})\cdot \mathcal{O}(\epsilon^{-1})$, as anticipated at the end of Section \ref{sec:metric}. To evaluate $\zeta_*$ using Eq.~\eqref{zeta}, we would therefore need to go to the next order in $\Psi_-$ and $\Phi_-$: we evade this by computing $\zeta_*$ directly. This can be achieved either by solving Eq.~\eqref{zetafirst} and inserting solution \eqref{psiminus} for $\Psi_-$, or by solving directly Eq.~\eqref{eq:zeta}. 

Let us start by solving Eq.~\eqref{eq:zeta}. The source term is 
\be
S_\zeta \equiv \frac{1}{\epsilon\, x^2\rho_\varphi}\left[-6\rho_\eb+x\frac{d\rho_\eb}{dx}+x\frac{d\pis}{dx}\right]\,,
\ee
involving not only the anisotropic stress but also the electromagnetic energy density and its derivative, which are
given in Eqs.~(\ref{e:Pemgen},\ref{e:rhoem'}). The electromagnetic energy density and the anisotropic stress have the same dependence on time and wavenumber, i.e. they go as $x^\alpha$. The integration of  Eq.~\eqref{eq:zeta} becomes then straightforward: the details of the solution are given in Appendix~\ref{app:zeta}. Identifying again the quantum operators $\hat\zeta$ and $\rho_\eb$ with their r.m.s. amplitude as done in Section \ref{sec:bardeeninflation}, one finally obtains the solution
\bea
\label{sol:zeta}
\zeta^-_{\rm inh}(x) &\simeq& \nonumber \\
&& \hspace{-1.2cm} \frac{H^2}{9m_P^2} \frac{1}{\epsilon} \Big[(\alpha-6)^2 C^2_\eb+2\al(\alpha-6)C_{\rho\Pi}+
     \alpha^2 \,C^2_\Pi \Big]^{1/2}  \nonumber \\
&\times& \left\{\begin{array}{cc} -\log\big(x/x_{\rm in}) &\mbox{if}~\al=0\\
x_{\rm in}^\alpha/\alpha & \mbox{if} ~\al\neq 0\end{array} \right.
\eea
where $x_{\rm in}\simeq 1$ denotes horizon exit (when the source starts to act). We have introduced the 
coefficient $C_{\rho\Pi}$ denoting the amplitude of the cross term arising from the correlation $\langle0|\rho_\eb^\dagger\Pi_S | 0\rangle$: the cross-correlation has the same spectral dependence, as $x^\alpha$, as the electromagnetic energy density and anisotropic stress.
We can define, analogously to Eq.~\eqref{hpis},
\be\label{Omem}
\Om_{\rm em}^-(k,\eta) \equiv \sqrt{\frac{k^3 P_\eb}{\rho^2_\vph}} \simeq  \frac{H^2}{3 m_P^2} C_{\rm em}(\ga) \, x^\al \, .
\ee
 It is interesting to note that in the scale-invariant case, $\al=0$, the anisotropic stress does not contribute 
to $\zeta^-_{\rm inh}$ to lowest order in the slow-roll expansion. Furthermore, only in the scale-invariant case there is a logarithmic build up of $\zeta^-_{\rm inh}$, whereas in all other cases $\zeta^-_{\rm inh}$ is constant on large scales: it is generated at horizon crossing and then stops growing. The Bardeen potentials $\Psi_-$ and $\Phi_-$, on the other hand, keep growing outside the horizon and soon become large: this renders longitudinal gauge badly adapted to the problem at hand. Conversely, in the comoving gauge, for example, all perturbation variables remain small during inflation (c.f. Appendix~\ref{app:zeta}). 

We could have solved for $\zeta^-_{\rm inh}$ by means of Eq.~\eqref{zetafirst} and inserting solution \eqref{psiminus} for $\Psi_-$. In this case, the source term in ~\eqref{zetafirst} contains also a contribution form the Poynting vector. However, this can be eliminated in favor of $\rho_\eb$ and $\Pi_S$ using the momentum conservation Eq.~\eqref{consmag} derived in Appendix ~\ref{app:zeta}, and the two approaches give finally the same result for $\zeta^-_{\rm inh}$. 

Neglecting the decaying mode $\Psi_1/x^3$ and using solution \eqref{sol:zeta} with $\zeta_*=\zeta^-_{\rm inh}(x_*)$, the Bardeen potentials in the radiation era are completely determined (c.f. Eq.~\eqref{psiplus}):  
\bea
\Psi_{+}(x) &=&\frac{2}{3}\left(1+\frac{1}{\epsilon} \right)b(k)+ \frac{2\zeta_*}{3}+\frac{3\hpisp}{x^2}\quad \mbox{and}  \nonumber\\
\Phi_{+}(x) &=& \frac{2}{3}\left(1+\frac{1}{\epsilon} \right)b(k) + \frac{2\zeta_*}{3} \; .
\label{psipluszeta}
\eea
Comparing these solutions with the ones obtained at large scales from a magnetic field generated causally during a primordial phase transition~\cite{cc}, we see that the only difference is the constant mode 
\bea 
&&\frac{2\zeta_*}{3} \sim \left(\frac{H_*}{m_P}\right)^2 \frac{1}{\epsilon}\left\{\begin{array}{cc} -\log\big(x_*) &\mbox{if}~\al=0\\
1/\alpha & \mbox{if} ~\al\neq 0\end{array} \right.
\label{zetastarmode}
\eea
(neglecting factors of order one). This constant mode is generated by the curvature perturbation at the end of inflation. 
In the case of a primordial phase transition, the curvature perturbation has no contribution from the electromagnetic field before the phase transition. 
The continuity of the curvature at the phase transition forces therefore $\zeta_*=0$.
During inflation, however, the curvature is dynamically generated by the electromagnetic source and at 
the end of inflation it transfers into the constant mode $\Psi_0$, adding a new term to the Bardeen potentials. 

In solutions \eqref{psipluszeta}, the mode sourced by the magnetic field anisotropic stress in the radiation era $\Om_\Pi^+$ would dominate at large scales. However, once neutrinos start free-streaming, this term is exactly compensated by the neutrino anisotropic stress, and leaves no visible effect in the CMB, see~\cite{cc}. The observationally relevant contribution from $\Om_\Pi^+$ is therefore given by the so-called `passive' mode, arising at next-to-leading order in the large-scale expansion \cite{shaw}. However, the passive mode is present also for a magnetic field generated by a causal process such as a phase transition; in the case of a magnetic field generated during inflation, one has to add the term proportional to $\zeta_*$ to the passive mode. The full solution for the Bardeen potential in the matter era from an inflationary magnetic field becomes therefore (in this discussion we neglect the contribution of the compensated mode, which is anyway subdominant with respect to the passive mode):
\be
\Psi_{+}(x>x_{\rm eq}) =\frac{2}{3}\left(1+\frac{1}{\epsilon} \right)b(k)+ \frac{2\zeta_*}{3}-\frac{3}{5}\, \hpisp \log\left(\frac{x_\nu}{x_*}\right)\,,
\label{Psitot}
\ee
where $x_{\rm eq}=k\eta_{\rm eq}$ with $\eta_{\rm eq}$ denoting the time of equality, and $x_{\nu}=k\eta_{\nu}$ with $\eta_{\nu}$ denoting the time of neutrino decoupling. The last term is the passive mode taken from Eq.~(6.10) of \cite{cc}, where we have set for the time of creation of the magnetic field $\eta=\eta_*$. Let us compare the amplitudes of the new constant mode $2\zeta_*/3$  and of the usual passive one. For sufficiently red spectra, $\alpha<3/2$, we can safely set $\hpis\simeq \hpisp$ and equally $\Om_{\rm em}^-\simeq \Om_{\rm em}^+$ (c.f. discussion at the end of Section \ref{Bradera}). The amplitude of the constant mode is therefore proportional to $(H_*/m_P)^2\sim \Om_{\rm em}^-/x_*^\alpha$, while the one of the passive mode is proportional to $\hpis\simeq \Om_{\rm em}^-$. Therefore, for a scale-invariant electromagnetic field with $\alpha=0$, the difference among the amplitude of the two modes is only due to the presence of the logarithms and of $1/\epsilon$ (c.f. Eq.~\eqref{zetastarmode}). The logarithm appearing in the term $2\zeta_*/3$ corresponds to the number of e-folds between horizon exit of a given scale $k\simeq 1/\eta_{\rm in}$ and the end of inflation, while the one appearing in the passive mode corresponds to the time interval between the end of inflation and neutrino free-streaming.  If, on the other hand, $0\neq \alpha<3/2$, the amplitude of the passive mode is significantly suppressed with respect to the constant inflationary one due to the extra factor $x_*^{\alpha}$.

Note also that the new term $2\zeta_*/3$ is of the same order of magnitude as the inflationary power spectrum
$P_b\simeq (H_*/m_P)^2/\ep$, up to the large log in the scale invariant case $\al=0$. However, coming from the 
square of a Gaussian field, it is genuinely non-Gaussian and may lead to a significant $f_{\rm NL}$, see Refs.~\cite{seery,barnaby}.

\section{ Discussion and Conclusions}
\label{sec:con}

In this paper we have computed the impact of primordial magnetic fields generated during inflation on the Bardeen potentials in the subsequent radiation era. We have solved the Bardeen equation both during inflation 
and in the radiation era. We have used the
inflationary solution as initial condition for the radiation era, by matching the solutions at the transition in such a way that the induced metric and the extrinsic curvature on the $\rho=$ constant hypersurface remain continuous through the transition. This matching procedure uniquely determines the solution after inflation. 

We have shown that the leading-order contribution to the Bardeen potential at large scales, $\Psi_-\simeq \hpis/(k\eta_*)^2$, is transferred entirely to the decaying mode in the radiation era. Consequently, the inflationary electromagnetic field contributes only at next-to-leading order, i.e. at order $\mathcal{O}(x^0)$.  A similar situation can appear in bouncing universes, like e.g. the ekpyrotic universe~\cite{web,CDC}: during the contracting phase, the growing mode can become very large. Only if this large mode is entirely transferred  to the decaying mode of the expanding Universe, do perturbations remain small; this full transfer can be achieved by the continuity of the curvature perturbation $\zeta$. Here the situation is different, because the large mode is due to a source. But we have found that even in this case, $\zeta$ remains small and its continuity is sufficient to guarantee that the large, superhorizon mode does not contribute to the constant $\Psi_0$. This shows that a gauge which exhibits large metric fluctuations, as the longitudinal one, can be very misleading: in the transition to the radiation era the Bardeen potential $\Phi$ jumps by a huge amount, while $\zeta$ remains continuous.

Performing the calculation at order $\mathcal{O}(x^0)$, we have found that the Bardeen potential in the radiation era contains a constant term proportional to the amplitude of the electromagnetic energy density at the end of inflation, which takes the form
\be \label{e:Psi+OmB}
\Psi_+^{\rm em}\propto \frac{\Om^-_{\rm em}(k,\eta_*)}{(k\eta_*)^\alpha}\frac{1}{\ep} \simeq 
\left(\frac{H_*}{m_P}\right)^2 \frac{1}{\ep} \, .
\ee
If the electromagnetic field generated during inflation is scale-invariant, this contribution is enhanced logarithmically by a 
factor of about $-\log(k\eta_* )$. Note that the origin of this constant term in $\Psi_+$ is quite different from the case of a magnetic field generated by a causal process, like, for example, at a phase transition in the early Universe. In this latter case, in fact, the matching conditions insure that the electromagnetic field perturbations are compensated on superhorizon scales. 

In addition to the above contribution to $\Psi_+$, there is the one from the anisotropic stress of the magnetic field 
during the radiation era $\propto \hpisp/x^2$, which is however compensated later on by neutrino free-streaming. As shown in \cite{cc} (see also~\cite{const}), the magnetic field anisotropic stress is not  compensated as long as the neutrinos are still coupled to radiation and do not free-stream. However, when neutrinos decouple, they develop an anisotropic stress that counter-balances the one of the magnetic field, removing the magnetic mode at large scales~\cite{Paoletti:2008ck}. Only the metric perturbations at next-to-leading order remain, which correspond to 
the so-called `passive' mode~\cite{shaw}.  
This compensation also takes place if the magnetic field is generated during inflation. 
However, as we have shown with this work, 
 in addition to the mode which later is compensated, $\propto {\hpisp}/{x^2}$, and to the passive mode, 
inflationary magnetic fields also lead to a constant mode as in Eq.~\eqref{e:Psi+OmB}.
 
Let us estimate the amplitude of the constant mode for the most interesting case of a scale-invariant magnetic field. In this case, the anisotropic stress does not contribute to the result since $\alpha=0$. The electromagnetic energy density is 
 \be
 \Om^-_{\rm em} \simeq\frac{\rho_B}{\bar\rho_{\rm rad}} \simeq \left(\frac{B}{4\mu{\rm G}}\right)^2 \simeq 
 \left(\frac{H_*}{m_P}\right)^2\,.
 \ee
 Furthermore, $H_*^2 \sim T^4_*/m_P^2\sim  m_P^2\, \Om^-_{\rm em} $ so that $T_* \sim m_P\,(\Om_{\rm em}^-)^{1/4}$. The largest observationally allowed magnetic fields have an amplitude $B\simeq 10^{-9}$G, leading to $ \Om_{\rm em}^-\simeq 10^{-7}$ and 
 $T_* \sim 10^{-2}m_P$.  For the conformal time at the transition we find
 $ \eta_* =\HH_*^{-1} = (a_*H_*)^{-1} = T_* \sqrt{\Om^-_{\rm em}}/(T_0\,m_P)$. 
 Setting $k_0=H_0 \simeq  T_0^2 / (m_P \sqrt{\Om_\ga})$,
 we obtain for the present Hubble scale $k_0=H_0$
 $$ x_{*0} = \eta_*k_0 = \frac{1}{\sqrt{\Om_\ga\Om_{\rm em}}}\frac{T_0T_*}{m_P^2} 
 \simeq 10^{-28} \, ;$$
 so that $-\log(x_{*0}) \sim 64$.  Therefore, the amplitude of the metric perturbation, $\Psi_+^{\rm em}
 \simeq -\Om^-_{\rm em}\log(x_*)/\ep$, at the wavenumber corresponding to the present Hubble scale is enhanced by the number of e-folds of inflation after the present Hubble scale has exited the horizon.

To estimate the effect of an inflationary magnetic field on the CMB at large scales, we can set very roughly $\Delta T/T\sim \Psi$: the full observationally relevant contribution is then given by the sum of the passive and the constant modes, as given in Eq.~\eqref{Psitot}. In the scale-invariant case $\alpha=0$, the logarithmic enhancement of the passive mode is $-3/5\,\log(T_*/T_\nu)\simeq -27$, and the one of the new inflationary contribution is of the order of the number of e-folds, therefore in total larger by about a factor $2/\ep$. We obtain then that the amplitude of the effect on the CMB is increased with respect to the naively expected amplitude $\simeq  \Om_{\rm em}^-$ by nearly 2 orders of magnitude due to the large logarithms. In other words, it might be possible to detect inflationary magnetic fields in the CMB down
to about $10^{-10}$G instead of the usual limit of $10^{-9}$G. The limit of about $10^{-9}$G was obtained, for example, in Refs.~\cite{CMBPMF} which, in order to be model-independent, only evaluate the effect of the compensated mode, corresponding to $\Psi_+^{\rm em}\sim \Om_{\rm em}^-$. Furthermore, contrary to the inflaton perturbations, the contribution from the magnetic field is inherently non- Gaussian as it is quadratic in the field amplitude. This leads to a nontrivial bispectrum, see Refs.~\cite{seery,barnaby}. 
 
It remains to be investigated whether it is consistent to ignore the gauge symmetry breaking which is necessary for the proposed mechanism to work, or whether it leads to other consequences which have to be taken into account.
For example the effects of a possible longitudinal gauge mode have to be studied. However, apart from fine-tuned examples like the
one proposed in Ref.~\cite{luk2},
at present this inflationary scenario seems to be the only viable candidate for primordial large-scale magnetic fields. If 
a causal generation mechanism is to be successful, a strong inverse cascade is needed in order
to move correlations from small to larger scales. This inverse cascade has to be more efficient than the 
one proposed e.g. in \cite{campanelli}, which has been shown to be insufficient~\cite{elisa}. The issue of the evolution of helical magnetic fields is still unsolved and interesting research in this direction is ongoing, see 
e.g.~\cite{Jurg}.

\acknowledgments{ 
It is a pleasure to thank Lukas Hollenstein, Anthony Lewis, Ian Moss and Raquel Ribeiro for useful discussions. 
CB is supported by a Herchel Smith Postdoctoral Fellowship and
by King's College Cambridge. RD and CB thank the CEA Saclay, where this work has been initiated, for hospitality.
RD is supported by the Swiss National Science Foundation.}

\appendix

\section{An evolution equation for the curvature perturbation}
\label{app:zeta}
\subsection{ In comoving gauge}
Here we derive an equation for the curvature perturbation $\zeta$ in the comoving gauge.
Since the curvature is gauge invariant, we can use the solution to compute the constant mode of the Bardeen potential $\Psi_0$, as shown in the main text.
We work in the gauge comoving with the total fluid, i.e. such that $q_i^{\rm tot}\equiv q_i^{\rm em}+q_i^{\varphi}=0$ and $B_i=0$.
The metric is (in Fourier space, we use the notation of~\cite{mybook})
\be
\begin{split}
ds^2=&a^2\bigg\{-(1+2A)d\eta^2\\
&+\big[(1+2H_L)\delta_{ij}+2H_TY_{ij} \big]dx^idx^j\bigg\}\; .
\end{split}
\ee
In this gauge the perturbed Einstein equations read
 \begin{align}
&\HH A -H'_L-\frac{H_T'}{3}=0\label{com0i}\\
&3\HH^2 A-3\HH H_L'-k^2\left(H_L+\frac{H_T}{3}\right)=\\
&4\pi G a^2\left\{\frac{\varphi_0'^2 A}{a^2}-\rho_\eb+\ik\left[\left(\HH+\frac{2\varphi_0''}{\varphi_0'} \right)\qj -\qj'\right] \right\} \nonumber \\
&-k^2A+H_T''+2\HH H_T'-k^2\left(H_L+\frac{H_T}{3}\right)=8\pi G a^2\pis \\
&\HH A'+(2\HH'+\HH^2)A-\frac{k^2 A}{3}-\frac{k^2}{3}\left(H_L+\frac{H_T}{3}\right)-2\HH H_L'\nonumber\\
&\!\!-H_L''=4\pi Ga^2\!\left\{\!-\frac{\varphi_0'^2 A}{a^2}+\frac{\rho_\eb}{3} + \ik\Big(3\HH\qj +\qj'\Big) \!\!\right\}
\end{align}
These equations can be rewritten in terms of the curvature perturbation $\zeta$ which is 
given by
\be
\label{zetacom}
\zeta=-H_L-\frac{H_T}{3}\; .
\ee
Using Eq.~\eqref{com0i} to express $A$ in terms of $\zeta$ and Eq.~\eqref{zetacom} to write $H_T$ in term of $H_L$ and $\zeta$, one finds the following set of coupled equations for $\zeta$ and $H_L$:
 \begin{align}
&-\left(2\HH+\frac{\HH'}{\HH} \right)\zeta'-3\HH H_L'+k^2\zeta=\label{zeta1}\\
&4\pi G a^2\left\{-\rho_\eb+\ik\left[\left(\HH+\frac{2\varphi_0''}{\varphi_0'} \right)\qj -\qj'\right] \right\} \nonumber\\
&3\zeta''+\left(6\HH-\frac{k^2}{\HH}\right)\zeta'-k^2\zeta+3H_L''+6\HH H_L'=-8\pi G a^2\pis \label{zeta2} \\
&\zeta''+\left(2\HH-\frac{k^2}{3\HH}\right)\zeta'-\frac{k^2}{3}\zeta+H_L''+2\HH H_L'=\label{zeta3}\\
&-4\pi Ga^2\left\{\frac{\rho_\eb}{3} + \ik\Big(3\HH\qj +\qj'\Big) \right\}\nonumber
\end{align}
Combining Eq.~\eqref{zeta2} with Eq.~\eqref{zeta3} one finds a conservation equation for the magnetic field
\be
\label{consmag}
\ik\qj'= -\frac{\rho_\eb}{3}+\frac{2\pis}{3}-3\HH\ik\qj\; .
\ee
Deriving Eq.~\eqref{zeta1} and combining it with Eqs.~\eqref{zeta2} and~\eqref{consmag} gives a second-order evolution equation for the curvature $\zeta$
\begin{align}
&\left(\HH-\frac{\HH'}{\HH} \right)\zeta''+\left[2\HH^2-2\HH'-\left(\frac{\HH'}{\HH}\right)'+\left(\frac{\HH'}{\HH}\right)^2\right]\zeta'\nonumber\\
&+k^2\left(\HH-\frac{\HH'}{\HH} \right)\zeta=\frac{8\pi G a^2}{3}\Bigg\{\left[-3\HH+\frac{\HH'}{\HH}+\frac{2\varphi_0''}{\varphi_0'} \right]\pis\nonumber\\
&-\pis'+\left[-6\HH+\frac{\HH'}{\HH}-\frac{\varphi_0''}{\varphi_0'} \right]\rho_\eb -\rho_\eb'\nonumber\\
&-3\ik\left[\frac{\HH'}{\HH}\frac{\varphi_0''}{\varphi_0'} -\left(\frac{\varphi_0''}{\varphi_0'}\right)'\right]\qj\Bigg\}\; .
\end{align}
At lowest order in slow-roll this equation becomes
\be
\label{eqzetacom}
\zeta''+2\HH \zeta'+k^2\zeta=-\frac{8\pi G a^2}{3\epsilon}\left\{6\rho_\eb+\frac{\rho_\eb'}{\HH}+\frac{\pis'}{\HH}\right\}\; .
\ee
The source term in Eq.~\eqref{eqzetacom} has no contribution proportional to $\pis/k^2$.
Hence as expected it remains small at all scales providing that the magnetic field is small.
The curvature seems therefore a more suitable variable than the Bardeen potentials to describe perturbations in the presence of a primordial magnetic field during inflation.
In terms of the variable $x=-k\eta$ Eq.~\eqref{eqzetacom} reads
\bea
\label{eqzetacomx}
\frac{d^2\zeta}{dx^2} -\frac{2}{x}\frac{d\zeta}{dx}+\zeta&=&
\frac{1}{\epsilon x^2\rho_\varphi}\left\{-6\rho_\eb+x\frac{d\rho_\eb}{dx}+x\frac{d\pis}{dx}\right\}\nonumber\\
&\equiv&S_\zeta \; .
\eea
As we argue in the main text, the time and $k$ dependence of $\rho_\eb$ are the same as for $\pis$, i.e. $\propto x^\alpha$. We can therefore rewrite $\rho_\eb$ and $\pis$ simply as
\be
\frac{\rho_\eb}{\rho_\varphi}=A_\eb x^\alpha\quad \mbox{and}\quad\frac{\pis}{\rho_\varphi}=A_\Pi x^\alpha\;,
\ee
where $A_\eb$ and $A_\Pi$ denote the amplitude of the quantum operators which do not contain any $x$ dependence and are therefore irrelevant for the integration of Eq.~\eqref{eqzetacomx}. With this, the source term for the curvature becomes
\be
S_\zeta=\frac{1}{\epsilon}\Big[(\alpha-6)A_\eb+\alpha A_\Pi \Big]x^{\alpha-2}\; .
\ee
The homogeneous solution to Eq.~\eqref{eqzetacomx} is
\be
\zeta=c_1 x^{3/2}J_{3/2}(x)+c_2 x^{3/2}J_{-3/2}(x)\;.
\ee
The solution of the inhomogeneous equation can be computed using the Wronskian method.
Integrating from $x_{\rm in}\simeq 1$ to $0<x\ll1$ one finds
\be
\hat\zeta_{\rm inh}=\frac{1}{3\epsilon}\Big[(\alpha-6)A_\eb+\alpha A_\Pi \Big] \left\{\begin{array}{cc} -\log\big(x/x_{\rm in}) &\mbox{if}~\al=0\\
x_{\rm in}^\alpha/\alpha & \mbox{if} ~\al\neq 0\end{array} \right. \,.
\ee
From the above solution, it appears that when one identifies the quantum operator $\hat\zeta_{\rm inh}$ with its r.m.s. amplitude, in general there will be also a cross term due to the correlation among the electromagnetic energy density and the anisotropic stress, $\langle0|A_\eb^\dagger A_\Pi | 0\rangle$ (c.f. Eq.~\eqref{sol:zeta} of the main text). 

\subsection{ In synchronous gauge}
The $\zeta$--equation of motion can also be obtained in synchronous gauge. There the metric is
\be
ds^2=a^2\left\{- d\eta^2+\big[(1+2H_L)\delta_{ij}+2H_TY_{ij} \big]dx^idx^j\right\}\; .
\ee
And the perturbed Einstein equations read
\begin{align}
&  H'_L + \frac{H_T'}{3}= 4\pi G\left(-\vph_0'\de\vph +\frac{iak^j}{k^2}q_{{\rm em }j} \right)\label{syn0i}\\
& 3\HH H_L' + k^2\left(H_L+\frac{H_T}{3}\right)=\\
& 4\pi G \left[\varphi_0'\de\vph'+V_{,\vph}a^2\de\vph +a^2\rho_\eb\right] \nonumber \\
&  H_T''+2\HH H_T'-k^2\left(H_L+\frac{H_T}{3}\right)=8\pi G a^2\pis \\
& -\left(H_L+\frac{H_T}{3}\right)''  -2\HH \left(H_L+\frac{H_T}{3}\right)' =\nonumber\\
& 4\pi G \left[\varphi_0'\de\vph'-V_{,\vph}a^2\de\vph +a^2p_\eb -\frac{2}{3}a^2\pis\right]
\end{align}
Combining these equations to eliminate $\de\vph$ and using that in this gauge
$$\zeta = -\frac{1}{\ep\HH}\left(H_L+\frac{H_T}{3}\right)' - \left(H_L+\frac{H_T}{3}\right) \,,$$
one can again derive Eq.~\eqref{eqzetacomx} for $\zeta$ in the slow-roll approximation.

\section{Computation of the anisotropic stress}
\label{app:stress}

Here we present a detailed computation of the anisotropic stress power spectrum.
First we need to compute the anisotropic stress $\Pi_S(\bk, \eta)$. 
As an example, we focus on one specific contribution, namely   
\be
\Pi_S(\bx,\eta)=-\frac{1}{2}\frac{f^2}{8\pi a^4}A'_i(\bx,\eta)A'_i(\bx,\eta) + \cdots \; .
\ee
Using the expansion in Eq.~\eqref{expansion} for $A_i$, the Fourier transform of the anisotropic stress reads
\begin{widetext}
\be
\begin{split}
&\hspace{0.5cm} \Pi_S(\bk,\eta)=
-\frac{f^2}{2a^4}\int d^3x e^{i\bk\cdot\bx}\int \frac{d^3k'}{(2\pi)^3}\int\frac{d^3q}{(2\pi)^3}
\sum_{\lambda\lambda'=1}^2\boe_{\la\, i}(\bk')\boe_{\la\, i}(\bq)\times \\
&\Bigg[b_\la(\bk')\left(\frac{\A(k',\eta)}{f} \right)'e^{i\bk'\cdot\bx}
+ b^\dagger_\la(\bk')\left(\frac{\A(k',\eta)}{f} \right)'e^{-i\bk'\cdot\bx}\Bigg]
\Bigg[b_{\la'}(\bq)\left(\frac{\A(q,\eta)}{f} \right)'e^{i\bq\cdot\bx}
+ b^\dagger_{\la'}(\bq)\left(\frac{\A(q,\eta)}{f} \right)'e^{-i\bq\cdot\bx}\Bigg] + \cdots\; .
\end{split}
\ee
This equation contains four different products of creation and annihilation operators.
For each product the integral over $d^3x$ can be performed, e.g.
\be
b_\la(\bk')b_{\la'}(\bq)\int d^3x e^{i\bx(\bk+\bk'+\bq)}=(2\pi)^3\delta(\bk+\bk'+\bq)
 b_\la(\bk')b_{\la'}(-\bk-\bk')\; ,
\ee
and similarly for the three other terms. With this the anisotropic stress becomes
\be
\begin{split}
&\Pi_S(\bk,\eta)=
-\frac{f^2}{2a^4}\int\frac{d^3k'}{(2\pi)^3}\sum_{\lambda\lambda'=1}^2\boe_{\la\, i}(\bk')\left(\frac{\A(k',\eta)}{f} \right)'\Bigg\{ 
\boe_{\la'\, i}(-\bk-\bk')\left(\frac{\A(|\bk+\bk'|,\eta)}{f} \right)'b_\la(\bk')b_{\la'}(-\bk-\bk')\\
&+ \boe_{\la'\, i}(\bk+\bk')\left(\frac{\A(|\bk+\bk'|,\eta)}{f} \right)'b_\la(\bk')b^\dagger_{\la'}(\bk+\bk')
+ \boe_{\la'\, i}(\bk'-\bk)\left(\frac{\A(|\bk'-\bk|,\eta)}{f} \right)'b^\dagger_\la(\bk')b_{\la'}(\bk'-\bk)\\
&+ \boe_{\la'\, i}(\bk-\bk')\left(\frac{\A(|\bk-\bk'|,\eta)}{f} \right)'b^\dagger_\la(\bk')b^\dagger_{\la'}(\bk-\bk')\Bigg\}
+ \cdots \; .
\end{split}
\label{pisapp}
\ee
Performing the same calculation for the other contributions in $\Pi_S$ we obtain Eq.~\eqref{pis}. 

We can then compute the spectrum of the anisotropic stress 
$\langle 0|\Pi_S^\dagger(\bq,\eta)\Pi_S(\bk,\eta) |0\rangle$. 
The only operators that contribute to the spectrum are $bb^\dagger bb^\dagger $ and $bbb^\dagger b^\dagger$.
The first operator corresponds to the case where one mode is created and destroyed and then a second mode is created an destroyed,
and the second operator corresponds to the case where two modes are created and then destroyed. 
The contribution from the first operator reads
\be
\langle 0|\Big(b_\al(\bq')b_{\al'}^\dagger(\bq+\bq')\Big)^\dagger b_\la(\bk')b_{\la'}^\dagger(\bk+\bk')|0 \rangle  
\sim \delta_{\la\la'}\delta_{\al\al'}\delta(\bk)\delta(\bq)\; .
\ee
This term does only contribute to the zero-mode $\bk=\bq=0$ and it has therefore no effect on the fluctuations, i.e. on the Bardeen potential.
The contribution from the second operator is
\be
\label{bbdagg}
\begin{split}
&\langle 0| b_{\al'}(\bq-\bq')b_\al(\bq') b_\la^\dagger(\bk')b_{\la'}^\dagger(\bk-\bk')|0 \rangle  \\
& =(2\pi)^3\delta_{\al\la}\delta(\bq'-\bk')\langle 0|b_{\al'}(\bq-\bq')b_{\la'}^\dagger(\bk-\bk')|0 \rangle
+\langle 0|b_{\al'}(\bq-\bq') b_\la^\dagger(\bk')b_\al(\bq')b_{\la'}^\dagger(\bk-\bk')|0 \rangle\\
& = (2\pi)^6\delta_{\al\la}\delta_{\al'\la'}\delta(\bq'-\bk')\delta(\bk-\bk'+\bq'-\bq) 
+(2\pi)^6\delta_{\al\la'}\delta_{\al'\la}\delta(\bk'+\bq'-\bq)\delta(\bq'+\bk'-\bk)\; ,
\end{split}
\ee
where for the first equality we have used the commutation relation between $b_\al$ and $b^\dagger_\la$.
The two terms in Eq.~\eqref{bbdagg} give both the same contribution to the power spectrum, which becomes

\be
\label{pipi}
\begin{split}
&\langle 0|\pis^\dagger(\bq,\eta)\pis(\bk,\eta)|0\rangle=\frac{9}{2a^8}\delta^3(\bq-\bk)\int d^3k' \Bigg\{
B^2_1(\bk,\bk')f^4\left|\left(\frac{\A(k',\eta)}{f} \right)'\right|^2\cdot\left|\left(\frac{\A(|\bk-\bk'|,\eta)}{f} \right)'\right|^2\\
&+B_2^2(\bk,\bk')\big|\A(k',\eta)\big|^2\cdot\big|\A(|\bk-\bk'|,\eta)\big|^2 
+ 2B_1(\bk,\bk')B_2(\bk,\bk')\A(k',\eta)\A(|\bk-\bk'|,\eta) \\
&\times f^2\left(\frac{\A^*(k',\eta)}{f} \right)'\left(\frac{\A^*(|\bk-\bk'|,\eta)}{f} \right)' \Bigg\}   \; ,
\end{split}
\ee
where
\be
\begin{split}
&B_1(\bk,\bk')=\sum_{\la,\la'=1}^{2}\left(\hat k^i\hat k^j-\frac{\delta^{ij}}{3}\right)\boe_{\la\, i}(\bk')\boe_{\la'\, j}(\bk-\bk')\,,\\
&B_2(\bk,\bk')=\sum_{\la,\la'=1}^{2}\left(\frac{\delta^{ij}}{3}-\hat k^i\hat k^j\right)\Big[\boe_{\la\, \ell}(\bk')k_i'-\boe_{\la\, i}(\bk')k_\ell'\Big]\cdot \Big[\boe_{\la'\, \ell}(\bk-\bk')(k_j-k_j')-\boe_{\la'\, j}(\bk-\bk')(k_\ell-k_\ell')\Big]\; .
\end{split}
\label{B1B2}
\ee
Equation~\eqref{pipi} contains integrals over the direction of $\bk'$ that are difficult to compute exactly. 
However, since we are mainly interested in the scaling of the anisotropic stress with $\eta$ and $k$, 
but not in its precise numerical value, we approximate these integrals with
\be
\frac{1}{4\pi}\int d\Omega_{\bk'}B_1^2(\bk,\bk')\left|\left(\frac{\A(|\bk-\bk'|,\eta)}{f} \right)'\right|^2
\simeq \si_1(\gamma)\left|\left(\frac{\A(|k-k'|,\eta)}{f} \right)'\right|^2\; ,
\ee
with $\si_1(\gamma)$ a constant which depends somewhat on the index $\gamma$ through the solution for $\A$ but which is always of order one since the terms in $B_1$ contain only unit vectors.
Similarly we approximate 
\be
\frac{1}{4\pi}\int d\Omega_{\bk'}B_2^2(\bk,\bk')\big|\A(|\bk-\bk'|,\eta)\big|^2
\simeq \si_2(\gamma)k'^2|k-k'|^2\big|\A(|k-k'|,\eta)\big|^2\; ,
\label{intB2}
\ee
and
\be
\frac{2}{4\pi}\int d\Omega_{\bk'}B_1(\bk,\bk')B_2(\bk,\bk')\A(|\bk-\bk'|,\eta)\left(\frac{\A^*(|\bk-\bk'|,\eta)}{f} \right)'
\simeq \si_3(\gamma)k'|k-k'|\A(|k-k'|,\eta)\big|\left(\frac{\A^*(|k-k'|,\eta)}{f} \right)'\; ,
\ee
with $\si_2(\gamma)$ and $\si_3(\gamma)$ two constants of order unity. 
With these approximations only the integral over the wavenumber $k'$ remains and we find for the anisotropic stress power spectrum
\be
\label{ppi}
\begin{split}
&P_\Pi(k,\eta)=\frac{9\cdot 4\pi}{2a^8}\int_0^{1/|\eta|} \frac {k'^2 dk'}{(2\pi)^3} 
\Bigg\{\si_1(\gamma)f^4\left|\left(\frac{\A(k',\eta)}{f} \right)'\right|^2\cdot\left|
  \left(\frac{\A(|k-k'|,\eta)}{f} \right)'\right|^2
+\si_2(\gamma)k'^2|k-k'|^2\big|\A(k',\eta)\big|^2 \\ & \hspace{1cm}\times\big|\A(|k-k'|,\eta)\big|^2 
+ \si_3(\gamma)k'|k-k'|\A(k',\eta)\A(|k-k'|,\eta) 
 f^2\left(\frac{\A^*(k',\eta)}{f} \right)'\left(\frac{\A^*(|k-k'|,\eta)}{f} \right)' \Bigg\} \; ,
\end{split}
\ee
where we neglect the contributions coming from $k'>1/|\eta|$ because for these the 
Bessel functions in $\A$, $J_\nu(k'|\eta|)$ and  $J_\nu(|(k-k')\eta|)$, lead to oscillations and the result is damped.
This equation can be written in terms of the power spectra for $\bB$, $\bE$ and $\bE \bB$, that have been calculated, for example,
in Refs.~\cite{yokoyama,subramanian} 
\bea
P_B &=& 4\pi \frac{k^2}{f^2a^4} |{\cal A}(k,\eta)|^2\; , \\
P_E &=& 4\pi\frac{1}{a^4} \left|\left(\frac{{\cal A}(k,\eta)}{f}\right)'\right|^2\; ,\\
P_{EB} &=& 4\pi\frac{k}{fa^4} \left(\frac{{\cal A}(k,\eta)}{f}\right)'{\cal A}^*(k,\eta)  \,.
\eea
With this we find
\be
\label{ppiP}
\begin{split}
P_\Pi(k,\eta)=18\pi f^4\int_0^{1/|\eta|} \frac {k'2dk'}{(2\pi)^3}\, 
\Bigg\{& \si_1(\gamma)P_E(k',\eta)P_E(|k-k'|,\eta)
+\si_2(\gamma)P_B(k',\eta)P_B(|k-k'|,\eta) \\
&+ \si_3(\gamma)P_{EB}(k',\eta)P_{EB}(|k-k'|,\eta)\Bigg\} \; .
\end{split}
\ee
\end{widetext}
The calculation of the electromagnetic energy density spectrum $P_{\rm em}(k,\eta)$
is analogous. The only difference is in the
angular dependence of the individual terms which yields different parameters $\si_1$, $\si_2$ and $\si_3$.

\section{The various contributions to the anisotropic stress power spectrum}
\label{app:PEandB}

Here we compute the spectrum of the anisotropic stress generated by the electric field $P_E$ and the cross term $P_{EB}$.
Let us start by the electric part. For $\gamma<-1/2$, we use the first line in Eq.~\eqref{PE}. 
Approximating the convolution with Eq.~\eqref{e:Iapprox} in Appendix~\ref{app:conv} we find
\be\label{e:PE1}
P_{\Pi^{(E)}} \simeq\frac{9|2c_1|^4\si_1}{4\pi^2(\ga+1/2)^4|\eta|^5a^8}\,\frac{1}{9+4\ga}\,.
\ee
For $\gamma>-1/2$ we take the second line in Eq.~\eqref{PE} which gives 
\be
\label{e:PE2}
P_{\Pi^{(E)} }\simeq\frac{9|c_2|^4\si_1(1-2\ga)^4}{4\pi^2|\eta|^5a^8}\left\{\begin{array}{l}
\frac{1}{5-4\ga}  \\  \mbox{if } -1/2<\ga<5/4 \\    \\
\frac{1-2\ga}{(4-2\ga)(5-4\ga)}\, x^{5-4\ga} \\  \mbox{if } 5/4<\ga<2 \end{array}\right. \,.
\ee
The computation of the cross term involves three cases. For $\ga<-1/2$, we use the first line in Eq.~\eqref{PEandB}
which gives
\be
\label{e:PEandB1}
P_{\Pi^{(EB)} }\simeq\frac{9|c_1|^4 \si_3}{4\pi^2(\ga+1/2)^2|\eta|^5a^8}\left\{\begin{array}{l}
\frac{2\ga+2}{(5+2\ga)(7+4\ga)}\, x^{7+4\ga} \\  \mbox{if } -2<\ga<-7/4\\    \\
\frac{1}{7+4\ga}  \\  \mbox{if } -7/4<\ga<-1/2 \,. \end{array}\right.
\ee
For $-1/2<\ga<1/2$, we take the second line in Eq.~\eqref{PEandB} and we obtain
\be\label{e:PEandB2}
P_{\Pi^{(EB)}} \simeq\frac{9|c_1|^2|c_2|^2\si_3(1-2\ga)^2}{4\pi^2 5|\eta|^5a^8}\,.
\ee
Finally for $1/2<\ga<2$ we use the third line in Eq.~\eqref{PEandB} which gives
\be
\label{e:PEandB3}
P_{\Pi^{(EB)} }\simeq\frac{9|c_2|^4\si_3(1-2\ga)^2}{4\pi^2|\eta|^5a^8}\left\{\begin{array}{l}
\frac{1}{7-4\ga}  \\  \mbox{if } 1/2<\ga<7/4 \\    \\
\frac{2-2\ga}{(5-2\ga)(7-4\ga)}\, x^{7-4\ga} \\  \mbox{if } 7/4<\ga<2 \end{array}\right. \,.
\ee
The result for $P_{\Pi^{(B)} }$ is given in Eq.~(\ref{PB2}).

\section{Convolution integrals}
\label{app:conv}
In the convolution $\int_0^{1/|\eta|} dk' k'^2 P_X(k')P_Y(|k-k'|)$ we usually have to integrate power laws.
Hence these integrals are of the form
\be
I(\al,\beta)(k) \equiv \int_0^{1/|\eta|} dk' k'^\al |k-k'|^\beta \,.
\ee
Integrals of this type are very common when dealing with primordial magnetic fields; the standard way of approximating them has been given first in \cite{astro-ph/9911040}. More refined analytical evaluations of integrals of this type are beyond the scope of this paper, see {\it e.g.} \cite{Paoletti:2008ck}.  We first note that these integrals 
require $\al+1>0$ in order to avoid an infrared singularity at $k'\ra 0$.
We then split the integral in its part $0<k'<k$ and $k<k'<1/|\eta|$. We use that $x=k|\eta| <1$, hence $0<k<1/|\eta|$.
In the first interval we approximate 
$ |k-k'| \sim k$ while in the second interval we set $ |k-k'| \sim k'$. With this approximation, which is 
certainly crude but retains the main characteristics of the behavior, we obtain
\be
\begin{split}
I(\al,\beta)(k) \simeq  \hspace{5cm}\\
 \frac{1}{\al+1}k^{\al+\beta+1} +  \frac{1}{\al+\beta+1}
\left(|\eta|^{-(\al+\beta+1)} - k^{\al+\beta+1}\right)\\ 
\simeq  \frac{1}{|\eta|^{\al+\beta+1}}\left\{ \begin{array}{ll} 
  \frac{1}{\al+\beta+1} & \mbox{if } \al+\beta+1 >0 \\
  \frac{\beta x^{\al+\beta+1}}{(\al+1)(\al+\beta+1)} & \mbox{if } \al+\beta+1 <0\,.
\end{array}\right. \label{e:Iapprox}
\end{split}
\ee
For the last $\simeq$ we have set $x=k|\eta|$ and we use $x<1$ to determine the dominant contribution.
Interestingly, such a convolution always either has a red spectrum, $\propto k^n,~ n=\al+\beta+1<0$ or it is 
white noise, $\propto k^0$. Blue spectra cannot be generated by a convolution. If small scales dominate, the integral is dominated by the upper cutoff which yields a white noise behavior.


\begin{thebibliography}{99}

\bibitem{filaments} See for example R. Beck, ASTRA 5, 43 (2009); \\
P.P. Kronberg et al, Astrophys. J. 676, 7079 (2008);\\
 M.L. Bernet et al, Nature 454, 302 (2008);\\
  L. Pentericci et al, A\&A Supp. Ser. 145, 121 (2000); \\
  M. Thierbach et al, A\&A 397, 53 (2003); \\
  F. Govoni and L. Feretti, Int. J. Mod. Phys. D13, 1549 (2004);\\
   C. Vogt and T.A. Ensslin, A\&A 434, 67 (2005);\\
    D. Guidetti et al, A\&A 483, 699 (2008);\\
    A. Bonafede et al, A\&A 513, A30 (2010); \\
    Y. Xu et al, Astrophys. J. 637, 19 (2006);\\
    I. Vovk, A. M. Taylor, D. Semikoz, A. Neronov, 
  Astrophys. Lett. {747}, L14 (2012).
    
\bibitem{voids} A. Neronov and I. Vovk, Science 328, 73 (2010); \\
F. Tavecchio et al, MNRAS Lett. 406, L70 (2010); \\
K. Dolag et al, Astrophys. J. Lett. 727, L4 (2011); \\
A.~M.~Taylor, I.~Vovk, A.~Neronov, Astron. \& Astrophys. {529},  A144 (2011) [arXiv:1101.0932 [astro-ph.HE]]; \\
K.~Takahashi, M.~Mori, K.~Ichiki and S.~Inoue, Astrophys. Lett. {744}, L7 (2012).

\bibitem{grasso} D. Grasso and H. Rubinstein, Phys. Rept. 348, 163 (2001); \\
M. Giovannini, Int. J. Mod. Phys. D 13, 391 (2004); \\
A.~Kandus, K.~E.~Kunze, C.~G.~Tsagas,
  Phys.\ Rept.\  {505 } (2011)  1-58.
  [arXiv:1007.3891 [astro-ph.CO]].
  
\bibitem{turner} M.S. Turner and L.M. Widrow, Phys. Rev. {D37}, 2743 (1988).

\bibitem{ratra} B. Ratra, Astrophys. J. Lett. {391}, L1 (1992).
  
\bibitem{yokoyama} J.~Martin, J.~'i.~Yokoyama,
  JCAP {0801 } (2008)  025
  [arXiv:0711.4307 [astro-ph]].
  
\bibitem{Campanelli:2008qp}
  L.~Campanelli, P.~Cea, G.~L.~Fogli and L.~Tedesco,
ÊÊPhys.\ Rev.\ D {77} (2008) 123002
ÊÊ[arXiv:0802.2630 [astro-ph]].
ÊÊ
  
\bibitem{bamba1}  K.~Bamba and J.~Yokoyama,
  Phys.\ Rev.\  D {69} (2004) 043507
  [arXiv:astro-ph/0310824].
  
\bibitem{bamba2}  K.~Bamba, N.~Ohta, S.~Tsujikawa,
  Phys.\ Rev.\  {D78 } (2008)  043524.
  [arXiv:0805.3862 [astro-ph]].
  
\bibitem{demozzi}
 V.~Demozzi, V.~Mukhanov, H.~Rubinstein,
  JCAP {0908 } (2009)  025.
  [arXiv:0907.1030 [astro-ph.CO]].
  
\bibitem{1109.4415} R. Caldwell, L. Motta and M. Kamionkowski, Phys. Rev. {D84}, 123525 (2011) [arXiv:1109.4415].

\bibitem{Durrer:2010mq} 
R. Durrer, L. Hollenstein and R.K. Jain,
     JCAP, {1103}, 037 (2011) [arXiv:1005.5322].

\bibitem{barnaby}  N. Barnaby, R. Namba and Marco Peloso, [arXiv:1202.1469]. 

\bibitem{muk_deruelle} 	N. Deruelle and V.F. Mukhanov, Phys. Rev. {D52}, 5549 (1995); \\
	N. Deruelle, D. Langlois and J.-P. Uzan, Phys. Rev. { D56}, 7608  (1997). 

\bibitem{mukhanov} V.F. Mukhanov, H.A. Feldman and R.H. Brandenberger, Phys. Rep. {215}, 203 (1992).

\bibitem{subramanian} 	K. Subramanian, Astron.Nachr. {331}, 110 (2010) [arXiv:0911.4771]. 
  
\bibitem{cc} C. Bonvin and C. Caprini, JCAP {1005}, 022 (2010). 

\bibitem{mybook}R. Durrer, {\em The Cosmic Microwave Background}, Cambridge 
         University Press (Cambridge, 2008).
         
\bibitem{suyama} T. Suyama and J. Yokoyama, [arXiv:1204.3976].         
      
        
\bibitem{AE} J.~Ahonen \& K.~Enqvist, {\it Phys. Lett.} {B382}, 40 (1996);\\
    G.~Baym \& H.~Heiselberg, {\it Phys. Rev.} {D56}, 5254 (1997).      
      
\bibitem{gr-qc/9504030}
  D.~Polarski and A.~A.~Starobinsky,
  Class.\ Quant.\ Grav.\ \ {13} (1996) 377
  [gr-qc/9504030].
    
\bibitem{gr-qc/9806066}
  C.~Kiefer, J.~Lesgourgues, D.~Polarski and A.~A.~Starobinsky,
  Class.\ Quant.\ Grav.\ \ {15} (1998) L67
  [gr-qc/9806066].

\bibitem{shaw} J.~R.~Shaw and A.~Lewis, Phys. Rev. {D81},  043517 (2010).  
     

    
\bibitem{seery} D. Seery, JCAP 0908:018 (2009) [arXiv:0810.1617 [astro-ph]].

  
\bibitem{web}R. Durrer, a web note (2001) [arXiv:hep-th/0112026].

\bibitem{CDC}C. Cartier, R. Durrer and E. Copeland, Phys. Rev. {D67},  103517 (2003)
            [arXiv:hep-th/0301198].

\bibitem{const} J. Adamek, R. Durrer, E. Fenu and M. Vonlanthen, JCAP06(2011)017 [arXiv:1102.5235].

\bibitem{Paoletti:2008ck}
  D.~Paoletti, F.~Finelli, F.~Paci,
  Mon.\ Not.\ Roy.\ Astron.\ Soc.\  {396 } (2009)  523-534.
  [arXiv:0811.0230 [astro-ph]];\\ 
  F. Finelli et al, Phys. Rev. D78, 023510 (2008).


     
\bibitem{CMBPMF}     
D. Paoletti and F. Finelli, Phys. Rev. D 83, 123533 (2011); D.G.~Yamazaki et al, Phys. Rev. D 81, 023008 (2010).

\bibitem{luk2}
C.T. Byrnes, L. Hollenstein, R.K. Jain and  F.R. Urban, JCAP 1203 (2012) 009 [arXiv:1111.2030].

            
\bibitem{campanelli}
L.~Campanelli,
\newblock Phys. Rev. Lett. {98}, 251302 (2007), [arXiv:0705.2308].


\bibitem{elisa}
C.~Caprini, R.~Durrer and E.~Fenu,
\newblock JCAP {0911}, 001 (2009)  [arXiv:0906.4976].

\bibitem{Jurg} A. Boyarski, O. Ruchayskiy and J. Fr\"ohlich,  Phys. Rev. Lett. 108, 031301 (2012) [arXiv:1109.3350].
  
\bibitem{astro-ph/9911040}
  R.~Durrer, P.~G.~Ferreira and T.~Kahniashvili,
  Phys.\ Rev.\ D\ {61} (2000) 043001
  [astro-ph/9911040].


         
\end{thebibliography}
\end{document}